\documentclass[longauth]{aaEC}
\usepackage{euclid}
\usepackage{lineno}
\usepackage{amsmath}
\usepackage{amssymb}
\usepackage{scalerel}
\usepackage[utf8]{inputenc}
\usepackage{graphicx}
\usepackage{txfonts}
\usepackage{natbib}
\usepackage{appendix}

\usepackage{hyperref} \hypersetup{ colorlinks=true, linkcolor=blue, filecolor=magenta, urlcolor=blue, citecolor=blue}
\usepackage{chngcntr}
\usepackage{etoolbox}
\usepackage{xcolor}
\usepackage{diagbox}
\usepackage{multicol}
\usepackage{comment}
\usepackage{siunitx}
\usepackage{makecell}

\newcommand{\VIS}{I_{\scriptscriptstyle\rm E} }

\newcommand{\mlim}{$M_{\mathrm{lim}}$}	
\newcommand{\pt}{$p_{\mathrm{t}}$}
\newcommand{\nsig}{$N_{\sigma}$}
\newcommand{\nsigtwo}{$N^{\mathrm{2D}}_{\sigma}$}
\newcommand{\nsigthree}{$N^{\mathrm{3D}}_{\sigma}$}
\newcommand{\disperse}{DisPerSE}
\newcommand{\dfil}{$d_{\mathrm{fil}}$}
\newcommand{\dskel}{$d_{\mathrm{skel}}$}
\newcommand{\dnode}{$d_{\mathrm{CP}}$}

\begin{document}
\title{\Euclid preparation}
\subtitle{Establishing the quality of the 2D reconstruction of the filaments of the cosmic web with DisPerSE using Euclid photometric redshifts}    

\newcommand{\orcid}[1]{\unskip\protect\href{https://orcid.org/#1}{\protect\includegraphics[width=8pt,clip]{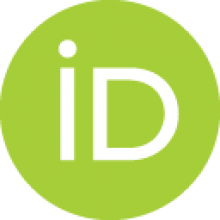}}}
\author{Euclid Collaboration: N.~Malavasi\orcid{0000-0001-9033-7958}\thanks{\email{malavasi@mpe.mpg.de}}\inst{\ref{aff1},\ref{aff2}}
\and F.~Sarron\orcid{0000-0001-8376-0360}\inst{\ref{aff3},\ref{aff4},\ref{aff5}}
\and U.~Kuchner\orcid{0000-0002-0035-5202}\inst{\ref{aff6}}
\and C.~Laigle\orcid{0009-0008-5926-818X}\inst{\ref{aff7}}
\and K.~Kraljic\orcid{0000-0001-6180-0245}\inst{\ref{aff8}}
\and P.~Jablonka\orcid{0000-0002-9655-1063}\inst{\ref{aff9}}
\and M.~Balogh\orcid{0000-0003-4849-9536}\inst{\ref{aff10},\ref{aff11}}
\and S.~Bardelli\orcid{0000-0002-8900-0298}\inst{\ref{aff12}}
\and M.~Bolzonella\orcid{0000-0003-3278-4607}\inst{\ref{aff12}}
\and J.~Brinchmann\orcid{0000-0003-4359-8797}\inst{\ref{aff13},\ref{aff14}}
\and G.~De~Lucia\orcid{0000-0002-6220-9104}\inst{\ref{aff15}}
\and F.~Fontanot\orcid{0000-0003-4744-0188}\inst{\ref{aff15},\ref{aff16}}
\and C.~Gouin\orcid{0000-0002-8837-9953}\inst{\ref{aff7}}
\and M.~Hirschmann\orcid{0000-0002-3301-3321}\inst{\ref{aff17}}
\and Y.~Kang\orcid{0009-0000-8588-7250}\inst{\ref{aff18}}
\and M.~Magliocchetti\orcid{0000-0001-9158-4838}\inst{\ref{aff19}}
\and T.~Moutard\orcid{0000-0002-3305-9901}\inst{\ref{aff20}}
\and J.~G.~Sorce\orcid{0000-0002-2307-2432}\inst{\ref{aff21},\ref{aff22},\ref{aff23}}
\and M.~Spinelli\orcid{0000-0003-0148-3254}\inst{\ref{aff24},\ref{aff15},\ref{aff25}}
\and L.~Wang\orcid{0000-0002-6736-9158}\inst{\ref{aff26},\ref{aff27}}
\and L.~Xie\orcid{0000-0003-3864-068X}\inst{\ref{aff28}}
\and A.~M.~C.~Le~Brun\orcid{0000-0002-0936-4594}\inst{\ref{aff29}}
\and E.~Tsaprazi\orcid{0000-0001-5082-4380}\inst{\ref{aff30}}
\and O.~Cucciati\orcid{0000-0002-9336-7551}\inst{\ref{aff12}}
\and G.~Zamorani\orcid{0000-0002-2318-301X}\inst{\ref{aff12}}
\and M.~De~Petris\orcid{0000-0001-7859-2139}\inst{\ref{aff31},\ref{aff32}}
\and E.~Bulbul\inst{\ref{aff1}}
\and R.~van~de~Weygaert\orcid{0000-0001-8379-1263}\inst{\ref{aff27}}
\and N.~Aghanim\orcid{0000-0002-6688-8992}\inst{\ref{aff22}}
\and A.~Amara\inst{\ref{aff33}}
\and S.~Andreon\orcid{0000-0002-2041-8784}\inst{\ref{aff34}}
\and N.~Auricchio\orcid{0000-0003-4444-8651}\inst{\ref{aff12}}
\and C.~Baccigalupi\orcid{0000-0002-8211-1630}\inst{\ref{aff16},\ref{aff15},\ref{aff35},\ref{aff36}}
\and M.~Baldi\orcid{0000-0003-4145-1943}\inst{\ref{aff37},\ref{aff12},\ref{aff38}}
\and A.~Biviano\orcid{0000-0002-0857-0732}\inst{\ref{aff15},\ref{aff16}}
\and E.~Branchini\orcid{0000-0002-0808-6908}\inst{\ref{aff39},\ref{aff40},\ref{aff34}}
\and M.~Brescia\orcid{0000-0001-9506-5680}\inst{\ref{aff41},\ref{aff42},\ref{aff43}}
\and S.~Camera\orcid{0000-0003-3399-3574}\inst{\ref{aff44},\ref{aff45},\ref{aff46}}
\and G.~Ca\~nas-Herrera\orcid{0000-0003-2796-2149}\inst{\ref{aff47},\ref{aff48}}
\and V.~Capobianco\orcid{0000-0002-3309-7692}\inst{\ref{aff46}}
\and C.~Carbone\orcid{0000-0003-0125-3563}\inst{\ref{aff49}}
\and J.~Carretero\orcid{0000-0002-3130-0204}\inst{\ref{aff50},\ref{aff51}}
\and S.~Casas\orcid{0000-0002-4751-5138}\inst{\ref{aff52},\ref{aff53}}
\and M.~Castellano\orcid{0000-0001-9875-8263}\inst{\ref{aff32}}
\and G.~Castignani\orcid{0000-0001-6831-0687}\inst{\ref{aff12}}
\and S.~Cavuoti\orcid{0000-0002-3787-4196}\inst{\ref{aff42},\ref{aff43}}
\and K.~C.~Chambers\orcid{0000-0001-6965-7789}\inst{\ref{aff54}}
\and C.~Colodro-Conde\inst{\ref{aff55}}
\and G.~Congedo\orcid{0000-0003-2508-0046}\inst{\ref{aff56}}
\and C.~J.~Conselice\orcid{0000-0003-1949-7638}\inst{\ref{aff57}}
\and L.~Conversi\orcid{0000-0002-6710-8476}\inst{\ref{aff58},\ref{aff20}}
\and Y.~Copin\orcid{0000-0002-5317-7518}\inst{\ref{aff59}}
\and F.~Courbin\orcid{0000-0003-0758-6510}\inst{\ref{aff60},\ref{aff61}}
\and H.~M.~Courtois\orcid{0000-0003-0509-1776}\inst{\ref{aff62}}
\and M.~Cropper\orcid{0000-0003-4571-9468}\inst{\ref{aff63}}
\and A.~Da~Silva\orcid{0000-0002-6385-1609}\inst{\ref{aff64},\ref{aff65}}
\and H.~Degaudenzi\orcid{0000-0002-5887-6799}\inst{\ref{aff18}}
\and S.~de~la~Torre\inst{\ref{aff66}}
\and A.~M.~Di~Giorgio\orcid{0000-0002-4767-2360}\inst{\ref{aff19}}
\and J.~Dinis\orcid{0000-0001-5075-1601}\inst{\ref{aff64},\ref{aff65}}
\and H.~Dole\orcid{0000-0002-9767-3839}\inst{\ref{aff22}}
\and F.~Dubath\orcid{0000-0002-6533-2810}\inst{\ref{aff18}}
\and C.~A.~J.~Duncan\orcid{0009-0003-3573-0791}\inst{\ref{aff57}}
\and X.~Dupac\inst{\ref{aff20}}
\and S.~Dusini\orcid{0000-0002-1128-0664}\inst{\ref{aff67}}
\and A.~Ealet\orcid{0000-0003-3070-014X}\inst{\ref{aff59}}
\and S.~Escoffier\orcid{0000-0002-2847-7498}\inst{\ref{aff68}}
\and M.~Farina\orcid{0000-0002-3089-7846}\inst{\ref{aff19}}
\and S.~Farrens\orcid{0000-0002-9594-9387}\inst{\ref{aff69}}
\and F.~Faustini\orcid{0000-0001-6274-5145}\inst{\ref{aff70},\ref{aff32}}
\and S.~Ferriol\inst{\ref{aff59}}
\and F.~Finelli\orcid{0000-0002-6694-3269}\inst{\ref{aff12},\ref{aff71}}
\and S.~Fotopoulou\orcid{0000-0002-9686-254X}\inst{\ref{aff72}}
\and M.~Frailis\orcid{0000-0002-7400-2135}\inst{\ref{aff15}}
\and E.~Franceschi\orcid{0000-0002-0585-6591}\inst{\ref{aff12}}
\and S.~Galeotta\orcid{0000-0002-3748-5115}\inst{\ref{aff15}}
\and K.~George\orcid{0000-0002-1734-8455}\inst{\ref{aff2}}
\and W.~Gillard\orcid{0000-0003-4744-9748}\inst{\ref{aff68}}
\and B.~Gillis\orcid{0000-0002-4478-1270}\inst{\ref{aff56}}
\and C.~Giocoli\orcid{0000-0002-9590-7961}\inst{\ref{aff12},\ref{aff38}}
\and P.~G\'omez-Alvarez\orcid{0000-0002-8594-5358}\inst{\ref{aff73},\ref{aff20}}
\and J.~Gracia-Carpio\inst{\ref{aff1}}
\and A.~Grazian\orcid{0000-0002-5688-0663}\inst{\ref{aff74}}
\and F.~Grupp\inst{\ref{aff1},\ref{aff2}}
\and L.~Guzzo\orcid{0000-0001-8264-5192}\inst{\ref{aff75},\ref{aff34}}
\and S.~V.~H.~Haugan\orcid{0000-0001-9648-7260}\inst{\ref{aff76}}
\and W.~Holmes\inst{\ref{aff77}}
\and I.~Hook\orcid{0000-0002-2960-978X}\inst{\ref{aff78}}
\and F.~Hormuth\inst{\ref{aff79}}
\and A.~Hornstrup\orcid{0000-0002-3363-0936}\inst{\ref{aff80},\ref{aff81}}
\and P.~Hudelot\inst{\ref{aff7}}
\and S.~Ili\'c\orcid{0000-0003-4285-9086}\inst{\ref{aff82},\ref{aff3}}
\and K.~Jahnke\orcid{0000-0003-3804-2137}\inst{\ref{aff83}}
\and M.~Jhabvala\inst{\ref{aff84}}
\and B.~Joachimi\orcid{0000-0001-7494-1303}\inst{\ref{aff85}}
\and E.~Keih\"anen\orcid{0000-0003-1804-7715}\inst{\ref{aff86}}
\and S.~Kermiche\orcid{0000-0002-0302-5735}\inst{\ref{aff68}}
\and A.~Kiessling\orcid{0000-0002-2590-1273}\inst{\ref{aff77}}
\and M.~Kilbinger\orcid{0000-0001-9513-7138}\inst{\ref{aff69}}
\and B.~Kubik\orcid{0009-0006-5823-4880}\inst{\ref{aff59}}
\and M.~K\"ummel\orcid{0000-0003-2791-2117}\inst{\ref{aff2}}
\and M.~Kunz\orcid{0000-0002-3052-7394}\inst{\ref{aff87}}
\and H.~Kurki-Suonio\orcid{0000-0002-4618-3063}\inst{\ref{aff88},\ref{aff89}}
\and S.~Ligori\orcid{0000-0003-4172-4606}\inst{\ref{aff46}}
\and P.~B.~Lilje\orcid{0000-0003-4324-7794}\inst{\ref{aff76}}
\and V.~Lindholm\orcid{0000-0003-2317-5471}\inst{\ref{aff88},\ref{aff89}}
\and I.~Lloro\orcid{0000-0001-5966-1434}\inst{\ref{aff90}}
\and G.~Mainetti\orcid{0000-0003-2384-2377}\inst{\ref{aff91}}
\and D.~Maino\inst{\ref{aff75},\ref{aff49},\ref{aff92}}
\and E.~Maiorano\orcid{0000-0003-2593-4355}\inst{\ref{aff12}}
\and O.~Mansutti\orcid{0000-0001-5758-4658}\inst{\ref{aff15}}
\and S.~Marcin\inst{\ref{aff93}}
\and O.~Marggraf\orcid{0000-0001-7242-3852}\inst{\ref{aff94}}
\and K.~Markovic\orcid{0000-0001-6764-073X}\inst{\ref{aff77}}
\and M.~Martinelli\orcid{0000-0002-6943-7732}\inst{\ref{aff32},\ref{aff95}}
\and N.~Martinet\orcid{0000-0003-2786-7790}\inst{\ref{aff66}}
\and F.~Marulli\orcid{0000-0002-8850-0303}\inst{\ref{aff96},\ref{aff12},\ref{aff38}}
\and R.~Massey\orcid{0000-0002-6085-3780}\inst{\ref{aff97}}
\and S.~Maurogordato\inst{\ref{aff24}}
\and H.~J.~McCracken\orcid{0000-0002-9489-7765}\inst{\ref{aff7}}
\and E.~Medinaceli\orcid{0000-0002-4040-7783}\inst{\ref{aff12}}
\and S.~Mei\orcid{0000-0002-2849-559X}\inst{\ref{aff98}}
\and M.~Melchior\inst{\ref{aff93}}
\and Y.~Mellier\inst{\ref{aff99},\ref{aff7}}
\and M.~Meneghetti\orcid{0000-0003-1225-7084}\inst{\ref{aff12},\ref{aff38}}
\and E.~Merlin\orcid{0000-0001-6870-8900}\inst{\ref{aff32}}
\and G.~Meylan\inst{\ref{aff9}}
\and A.~Mora\orcid{0000-0002-1922-8529}\inst{\ref{aff100}}
\and M.~Moresco\orcid{0000-0002-7616-7136}\inst{\ref{aff96},\ref{aff12}}
\and L.~Moscardini\orcid{0000-0002-3473-6716}\inst{\ref{aff96},\ref{aff12},\ref{aff38}}
\and S.~Mourre\inst{\ref{aff24},\ref{aff101}}
\and C.~Neissner\orcid{0000-0001-8524-4968}\inst{\ref{aff102},\ref{aff51}}
\and S.-M.~Niemi\inst{\ref{aff47}}
\and C.~Padilla\orcid{0000-0001-7951-0166}\inst{\ref{aff102}}
\and S.~Paltani\orcid{0000-0002-8108-9179}\inst{\ref{aff18}}
\and F.~Pasian\orcid{0000-0002-4869-3227}\inst{\ref{aff15}}
\and K.~Pedersen\inst{\ref{aff103}}
\and V.~Pettorino\inst{\ref{aff47}}
\and S.~Pires\orcid{0000-0002-0249-2104}\inst{\ref{aff69}}
\and G.~Polenta\orcid{0000-0003-4067-9196}\inst{\ref{aff70}}
\and M.~Poncet\inst{\ref{aff104}}
\and L.~A.~Popa\inst{\ref{aff105}}
\and L.~Pozzetti\orcid{0000-0001-7085-0412}\inst{\ref{aff12}}
\and F.~Raison\orcid{0000-0002-7819-6918}\inst{\ref{aff1}}
\and R.~Rebolo\inst{\ref{aff55},\ref{aff106},\ref{aff107}}
\and A.~Renzi\orcid{0000-0001-9856-1970}\inst{\ref{aff108},\ref{aff67}}
\and J.~Rhodes\orcid{0000-0002-4485-8549}\inst{\ref{aff77}}
\and G.~Riccio\inst{\ref{aff42}}
\and E.~Romelli\orcid{0000-0003-3069-9222}\inst{\ref{aff15}}
\and M.~Roncarelli\orcid{0000-0001-9587-7822}\inst{\ref{aff12}}
\and E.~Rossetti\orcid{0000-0003-0238-4047}\inst{\ref{aff37}}
\and R.~Saglia\orcid{0000-0003-0378-7032}\inst{\ref{aff2},\ref{aff1}}
\and Z.~Sakr\orcid{0000-0002-4823-3757}\inst{\ref{aff109},\ref{aff3},\ref{aff110}}
\and A.~G.~S\'anchez\orcid{0000-0003-1198-831X}\inst{\ref{aff1}}
\and D.~Sapone\orcid{0000-0001-7089-4503}\inst{\ref{aff111}}
\and B.~Sartoris\orcid{0000-0003-1337-5269}\inst{\ref{aff2},\ref{aff15}}
\and J.~A.~Schewtschenko\orcid{0000-0002-4913-6393}\inst{\ref{aff56}}
\and M.~Schirmer\orcid{0000-0003-2568-9994}\inst{\ref{aff83}}
\and P.~Schneider\orcid{0000-0001-8561-2679}\inst{\ref{aff94}}
\and T.~Schrabback\orcid{0000-0002-6987-7834}\inst{\ref{aff112}}
\and M.~Scodeggio\inst{\ref{aff49}}
\and A.~Secroun\orcid{0000-0003-0505-3710}\inst{\ref{aff68}}
\and E.~Sefusatti\orcid{0000-0003-0473-1567}\inst{\ref{aff15},\ref{aff16},\ref{aff35}}
\and G.~Seidel\orcid{0000-0003-2907-353X}\inst{\ref{aff83}}
\and S.~Serrano\orcid{0000-0002-0211-2861}\inst{\ref{aff113},\ref{aff114},\ref{aff115}}
\and P.~Simon\inst{\ref{aff94}}
\and C.~Sirignano\orcid{0000-0002-0995-7146}\inst{\ref{aff108},\ref{aff67}}
\and G.~Sirri\orcid{0000-0003-2626-2853}\inst{\ref{aff38}}
\and A.~Spurio~Mancini\orcid{0000-0001-5698-0990}\inst{\ref{aff116},\ref{aff63}}
\and L.~Stanco\orcid{0000-0002-9706-5104}\inst{\ref{aff67}}
\and J.~Steinwagner\orcid{0000-0001-7443-1047}\inst{\ref{aff1}}
\and P.~Tallada-Cresp\'{i}\orcid{0000-0002-1336-8328}\inst{\ref{aff50},\ref{aff51}}
\and A.~N.~Taylor\inst{\ref{aff56}}
\and I.~Tereno\inst{\ref{aff64},\ref{aff117}}
\and S.~Toft\orcid{0000-0003-3631-7176}\inst{\ref{aff118},\ref{aff119}}
\and R.~Toledo-Moreo\orcid{0000-0002-2997-4859}\inst{\ref{aff120}}
\and F.~Torradeflot\orcid{0000-0003-1160-1517}\inst{\ref{aff51},\ref{aff50}}
\and A.~Tsyganov\inst{\ref{aff121}}
\and I.~Tutusaus\orcid{0000-0002-3199-0399}\inst{\ref{aff3}}
\and L.~Valenziano\orcid{0000-0002-1170-0104}\inst{\ref{aff12},\ref{aff71}}
\and J.~Valiviita\orcid{0000-0001-6225-3693}\inst{\ref{aff88},\ref{aff89}}
\and T.~Vassallo\orcid{0000-0001-6512-6358}\inst{\ref{aff2},\ref{aff15}}
\and G.~Verdoes~Kleijn\orcid{0000-0001-5803-2580}\inst{\ref{aff27}}
\and A.~Veropalumbo\orcid{0000-0003-2387-1194}\inst{\ref{aff34},\ref{aff40},\ref{aff122}}
\and Y.~Wang\orcid{0000-0002-4749-2984}\inst{\ref{aff123}}
\and J.~Weller\orcid{0000-0002-8282-2010}\inst{\ref{aff2},\ref{aff1}}
\and A.~Zacchei\orcid{0000-0003-0396-1192}\inst{\ref{aff15},\ref{aff16}}
\and I.~A.~Zinchenko\orcid{0000-0002-2944-2449}\inst{\ref{aff2}}
\and E.~Zucca\orcid{0000-0002-5845-8132}\inst{\ref{aff12}}
\and E.~Bozzo\orcid{0000-0002-8201-1525}\inst{\ref{aff18}}
\and C.~Burigana\orcid{0000-0002-3005-5796}\inst{\ref{aff124},\ref{aff71}}
\and M.~Calabrese\orcid{0000-0002-2637-2422}\inst{\ref{aff125},\ref{aff49}}
\and D.~Di~Ferdinando\inst{\ref{aff38}}
\and J.~A.~Escartin~Vigo\inst{\ref{aff1}}
\and L.~Gabarra\orcid{0000-0002-8486-8856}\inst{\ref{aff126}}
\and S.~Matthew\orcid{0000-0001-8448-1697}\inst{\ref{aff56}}
\and N.~Mauri\orcid{0000-0001-8196-1548}\inst{\ref{aff127},\ref{aff38}}
\and A.~Pezzotta\orcid{0000-0003-0726-2268}\inst{\ref{aff1}}
\and M.~P\"ontinen\orcid{0000-0001-5442-2530}\inst{\ref{aff88}}
\and C.~Porciani\orcid{0000-0002-7797-2508}\inst{\ref{aff94}}
\and V.~Scottez\inst{\ref{aff99},\ref{aff128}}
\and M.~Tenti\orcid{0000-0002-4254-5901}\inst{\ref{aff38}}
\and M.~Viel\orcid{0000-0002-2642-5707}\inst{\ref{aff16},\ref{aff15},\ref{aff36},\ref{aff35},\ref{aff129}}
\and M.~Wiesmann\orcid{0009-0000-8199-5860}\inst{\ref{aff76}}
\and Y.~Akrami\orcid{0000-0002-2407-7956}\inst{\ref{aff130},\ref{aff131}}
\and V.~Allevato\orcid{0000-0001-7232-5152}\inst{\ref{aff42}}
\and I.~T.~Andika\orcid{0000-0001-6102-9526}\inst{\ref{aff132},\ref{aff133}}
\and S.~Anselmi\orcid{0000-0002-3579-9583}\inst{\ref{aff67},\ref{aff108},\ref{aff29}}
\and M.~Archidiacono\orcid{0000-0003-4952-9012}\inst{\ref{aff75},\ref{aff92}}
\and F.~Atrio-Barandela\orcid{0000-0002-2130-2513}\inst{\ref{aff134}}
\and A.~Balaguera-Antolinez\orcid{0000-0001-5028-3035}\inst{\ref{aff55},\ref{aff107}}
\and M.~Ballardini\orcid{0000-0003-4481-3559}\inst{\ref{aff135},\ref{aff12},\ref{aff136}}
\and C.~Benoist\inst{\ref{aff24}}
\and D.~Bertacca\orcid{0000-0002-2490-7139}\inst{\ref{aff108},\ref{aff74},\ref{aff67}}
\and M.~Bethermin\orcid{0000-0002-3915-2015}\inst{\ref{aff8}}
\and A.~Blanchard\orcid{0000-0001-8555-9003}\inst{\ref{aff3}}
\and L.~Blot\orcid{0000-0002-9622-7167}\inst{\ref{aff137},\ref{aff29}}
\and H.~B\"ohringer\orcid{0000-0001-8241-4204}\inst{\ref{aff1},\ref{aff138},\ref{aff139}}
\and S.~Borgani\orcid{0000-0001-6151-6439}\inst{\ref{aff140},\ref{aff16},\ref{aff15},\ref{aff35},\ref{aff129}}
\and M.~L.~Brown\orcid{0000-0002-0370-8077}\inst{\ref{aff57}}
\and S.~Bruton\orcid{0000-0002-6503-5218}\inst{\ref{aff141}}
\and R.~Cabanac\orcid{0000-0001-6679-2600}\inst{\ref{aff3}}
\and A.~Calabro\orcid{0000-0003-2536-1614}\inst{\ref{aff32}}
\and B.~Camacho~Quevedo\orcid{0000-0002-8789-4232}\inst{\ref{aff113},\ref{aff115}}
\and A.~Cappi\inst{\ref{aff12},\ref{aff24}}
\and F.~Caro\inst{\ref{aff32}}
\and C.~S.~Carvalho\inst{\ref{aff117}}
\and T.~Castro\orcid{0000-0002-6292-3228}\inst{\ref{aff15},\ref{aff35},\ref{aff16},\ref{aff129}}
\and F.~Cogato\orcid{0000-0003-4632-6113}\inst{\ref{aff96},\ref{aff12}}
\and S.~Contarini\orcid{0000-0002-9843-723X}\inst{\ref{aff1}}
\and T.~Contini\orcid{0000-0003-0275-938X}\inst{\ref{aff3}}
\and A.~R.~Cooray\orcid{0000-0002-3892-0190}\inst{\ref{aff142}}
\and S.~Davini\orcid{0000-0003-3269-1718}\inst{\ref{aff40}}
\and F.~De~Paolis\orcid{0000-0001-6460-7563}\inst{\ref{aff143},\ref{aff144},\ref{aff145}}
\and G.~Desprez\orcid{0000-0001-8325-1742}\inst{\ref{aff27}}
\and A.~D\'iaz-S\'anchez\orcid{0000-0003-0748-4768}\inst{\ref{aff146}}
\and J.~J.~Diaz\inst{\ref{aff147}}
\and S.~Di~Domizio\orcid{0000-0003-2863-5895}\inst{\ref{aff39},\ref{aff40}}
\and J.~M.~Diego\orcid{0000-0001-9065-3926}\inst{\ref{aff148}}
\and A.~G.~Ferrari\orcid{0009-0005-5266-4110}\inst{\ref{aff38}}
\and P.~G.~Ferreira\orcid{0000-0002-3021-2851}\inst{\ref{aff126}}
\and A.~Finoguenov\orcid{0000-0002-4606-5403}\inst{\ref{aff88}}
\and A.~Fontana\orcid{0000-0003-3820-2823}\inst{\ref{aff32}}
\and K.~Ganga\orcid{0000-0001-8159-8208}\inst{\ref{aff98}}
\and J.~Garc\'ia-Bellido\orcid{0000-0002-9370-8360}\inst{\ref{aff130}}
\and T.~Gasparetto\orcid{0000-0002-7913-4866}\inst{\ref{aff15}}
\and E.~Gaztanaga\orcid{0000-0001-9632-0815}\inst{\ref{aff115},\ref{aff113},\ref{aff53}}
\and F.~Giacomini\orcid{0000-0002-3129-2814}\inst{\ref{aff38}}
\and F.~Gianotti\orcid{0000-0003-4666-119X}\inst{\ref{aff12}}
\and G.~Gozaliasl\orcid{0000-0002-0236-919X}\inst{\ref{aff149}}
\and A.~Gregorio\orcid{0000-0003-4028-8785}\inst{\ref{aff140},\ref{aff15},\ref{aff35}}
\and M.~Guidi\orcid{0000-0001-9408-1101}\inst{\ref{aff37},\ref{aff12}}
\and C.~M.~Gutierrez\orcid{0000-0001-7854-783X}\inst{\ref{aff150}}
\and A.~Hall\orcid{0000-0002-3139-8651}\inst{\ref{aff56}}
\and S.~Hemmati\orcid{0000-0003-2226-5395}\inst{\ref{aff151}}
\and C.~Hern\'andez-Monteagudo\orcid{0000-0001-5471-9166}\inst{\ref{aff107},\ref{aff55}}
\and H.~Hildebrandt\orcid{0000-0002-9814-3338}\inst{\ref{aff152}}
\and J.~Hjorth\orcid{0000-0002-4571-2306}\inst{\ref{aff103}}
\and A.~Jimenez~Mu\~noz\orcid{0009-0004-5252-185X}\inst{\ref{aff153}}
\and J.~J.~E.~Kajava\orcid{0000-0002-3010-8333}\inst{\ref{aff154},\ref{aff155}}
\and V.~Kansal\orcid{0000-0002-4008-6078}\inst{\ref{aff156},\ref{aff157}}
\and D.~Karagiannis\orcid{0000-0002-4927-0816}\inst{\ref{aff135},\ref{aff25}}
\and C.~C.~Kirkpatrick\inst{\ref{aff86}}
\and S.~Kruk\orcid{0000-0001-8010-8879}\inst{\ref{aff20}}
\and M.~Lattanzi\orcid{0000-0003-1059-2532}\inst{\ref{aff136}}
\and S.~Lee\orcid{0000-0002-8289-740X}\inst{\ref{aff77}}
\and J.~Le~Graet\orcid{0000-0001-6523-7971}\inst{\ref{aff68}}
\and L.~Legrand\orcid{0000-0003-0610-5252}\inst{\ref{aff158},\ref{aff159}}
\and M.~Lembo\orcid{0000-0002-5271-5070}\inst{\ref{aff135},\ref{aff136}}
\and J.~Lesgourgues\orcid{0000-0001-7627-353X}\inst{\ref{aff52}}
\and T.~I.~Liaudat\orcid{0000-0002-9104-314X}\inst{\ref{aff160}}
\and S.~J.~Liu\orcid{0000-0001-7680-2139}\inst{\ref{aff19}}
\and A.~Loureiro\orcid{0000-0002-4371-0876}\inst{\ref{aff161},\ref{aff30}}
\and J.~Macias-Perez\orcid{0000-0002-5385-2763}\inst{\ref{aff153}}
\and G.~Maggio\orcid{0000-0003-4020-4836}\inst{\ref{aff15}}
\and F.~Mannucci\orcid{0000-0002-4803-2381}\inst{\ref{aff162}}
\and R.~Maoli\orcid{0000-0002-6065-3025}\inst{\ref{aff31},\ref{aff32}}
\and J.~Mart\'{i}n-Fleitas\orcid{0000-0002-8594-569X}\inst{\ref{aff100}}
\and C.~J.~A.~P.~Martins\orcid{0000-0002-4886-9261}\inst{\ref{aff163},\ref{aff13}}
\and L.~Maurin\orcid{0000-0002-8406-0857}\inst{\ref{aff22}}
\and R.~B.~Metcalf\orcid{0000-0003-3167-2574}\inst{\ref{aff96},\ref{aff12}}
\and M.~Miluzio\inst{\ref{aff20},\ref{aff164}}
\and P.~Monaco\orcid{0000-0003-2083-7564}\inst{\ref{aff140},\ref{aff15},\ref{aff35},\ref{aff16}}
\and C.~Moretti\orcid{0000-0003-3314-8936}\inst{\ref{aff36},\ref{aff129},\ref{aff15},\ref{aff16},\ref{aff35}}
\and G.~Morgante\inst{\ref{aff12}}
\and K.~Naidoo\orcid{0000-0002-9182-1802}\inst{\ref{aff53}}
\and A.~Navarro-Alsina\orcid{0000-0002-3173-2592}\inst{\ref{aff94}}
\and S.~Nesseris\orcid{0000-0002-0567-0324}\inst{\ref{aff130}}
\and K.~Paterson\orcid{0000-0001-8340-3486}\inst{\ref{aff83}}
\and L.~Patrizii\inst{\ref{aff38}}
\and A.~Pisani\orcid{0000-0002-6146-4437}\inst{\ref{aff68},\ref{aff165}}
\and V.~Popa\orcid{0000-0002-9118-8330}\inst{\ref{aff105}}
\and D.~Potter\orcid{0000-0002-0757-5195}\inst{\ref{aff166}}
\and I.~Risso\orcid{0000-0003-2525-7761}\inst{\ref{aff167}}
\and P.-F.~Rocci\inst{\ref{aff22}}
\and M.~Sahl\'en\orcid{0000-0003-0973-4804}\inst{\ref{aff168}}
\and E.~Sarpa\orcid{0000-0002-1256-655X}\inst{\ref{aff36},\ref{aff129},\ref{aff35}}
\and A.~Schneider\orcid{0000-0001-7055-8104}\inst{\ref{aff166}}
\and D.~Sciotti\orcid{0009-0008-4519-2620}\inst{\ref{aff32},\ref{aff95}}
\and E.~Sellentin\inst{\ref{aff169},\ref{aff170}}
\and M.~Sereno\orcid{0000-0003-0302-0325}\inst{\ref{aff12},\ref{aff38}}
\and A.~Silvestri\orcid{0000-0001-6904-5061}\inst{\ref{aff48}}
\and L.~C.~Smith\orcid{0000-0002-3259-2771}\inst{\ref{aff171}}
\and S.~A.~Stanford\orcid{0000-0003-0122-0841}\inst{\ref{aff172}}
\and K.~Tanidis\inst{\ref{aff126}}
\and C.~Tao\orcid{0000-0001-7961-8177}\inst{\ref{aff68}}
\and G.~Testera\inst{\ref{aff40}}
\and R.~Teyssier\orcid{0000-0001-7689-0933}\inst{\ref{aff165}}
\and S.~Tosi\orcid{0000-0002-7275-9193}\inst{\ref{aff39},\ref{aff40}}
\and A.~Troja\orcid{0000-0003-0239-4595}\inst{\ref{aff108},\ref{aff67}}
\and M.~Tucci\inst{\ref{aff18}}
\and C.~Valieri\inst{\ref{aff38}}
\and D.~Vergani\orcid{0000-0003-0898-2216}\inst{\ref{aff12}}
\and G.~Verza\orcid{0000-0002-1886-8348}\inst{\ref{aff173}}
\and N.~A.~Walton\orcid{0000-0003-3983-8778}\inst{\ref{aff171}}}
										   
\institute{Max Planck Institute for Extraterrestrial Physics, Giessenbachstr. 1, 85748 Garching, Germany\label{aff1}
\and
Universit\"ats-Sternwarte M\"unchen, Fakult\"at f\"ur Physik, Ludwig-Maximilians-Universit\"at M\"unchen, Scheinerstrasse 1, 81679 M\"unchen, Germany\label{aff2}
\and
Institut de Recherche en Astrophysique et Plan\'etologie (IRAP), Universit\'e de Toulouse, CNRS, UPS, CNES, 14 Av. Edouard Belin, 31400 Toulouse, France\label{aff3}
\and
Institut de Recherche en Informatique de Toulouse (IRIT), Universit\'e de Toulouse, CNRS, Toulouse INP, UT3, 31062 Toulouse, France\label{aff4}
\and
Laboratoire MCD, Centre de Biologie Int\'egrative (CBI), Universit\'e de Toulouse, CNRS, UT3, 31062 Toulouse, France\label{aff5}
\and
University of Nottingham, University Park, Nottingham NG7 2RD, UK\label{aff6}
\and
Institut d'Astrophysique de Paris, UMR 7095, CNRS, and Sorbonne Universit\'e, 98 bis boulevard Arago, 75014 Paris, France\label{aff7}
\and
Universit\'e de Strasbourg, CNRS, Observatoire astronomique de Strasbourg, UMR 7550, 67000 Strasbourg, France\label{aff8}
\and
Institute of Physics, Laboratory of Astrophysics, Ecole Polytechnique F\'ed\'erale de Lausanne (EPFL), Observatoire de Sauverny, 1290 Versoix, Switzerland\label{aff9}
\and
Department of Physics and Astronomy, University of Waterloo, Waterloo, Ontario N2L 3G1, Canada\label{aff10}
\and
Waterloo Centre for Astrophysics, University of Waterloo, Waterloo, Ontario N2L 3G1, Canada\label{aff11}
\and
INAF-Osservatorio di Astrofisica e Scienza dello Spazio di Bologna, Via Piero Gobetti 93/3, 40129 Bologna, Italy\label{aff12}
\and
Instituto de Astrof\'isica e Ci\^encias do Espa\c{c}o, Universidade do Porto, CAUP, Rua das Estrelas, PT4150-762 Porto, Portugal\label{aff13}
\and
Faculdade de Ci\^encias da Universidade do Porto, Rua do Campo de Alegre, 4150-007 Porto, Portugal\label{aff14}
\and
INAF-Osservatorio Astronomico di Trieste, Via G. B. Tiepolo 11, 34143 Trieste, Italy\label{aff15}
\and
IFPU, Institute for Fundamental Physics of the Universe, via Beirut 2, 34151 Trieste, Italy\label{aff16}
\and
Institute of Physics, Laboratory for Galaxy Evolution, Ecole Polytechnique F\'ed\'erale de Lausanne, Observatoire de Sauverny, CH-1290 Versoix, Switzerland\label{aff17}
\and
Department of Astronomy, University of Geneva, ch. d'Ecogia 16, 1290 Versoix, Switzerland\label{aff18}
\and
INAF-Istituto di Astrofisica e Planetologia Spaziali, via del Fosso del Cavaliere, 100, 00100 Roma, Italy\label{aff19}
\and
ESAC/ESA, Camino Bajo del Castillo, s/n., Urb. Villafranca del Castillo, 28692 Villanueva de la Ca\~nada, Madrid, Spain\label{aff20}
\and
Univ. Lille, CNRS, Centrale Lille, UMR 9189 CRIStAL, 59000 Lille, France\label{aff21}
\and
Universit\'e Paris-Saclay, CNRS, Institut d'astrophysique spatiale, 91405, Orsay, France\label{aff22}
\and
Leibniz-Institut f\"{u}r Astrophysik (AIP), An der Sternwarte 16, 14482 Potsdam, Germany\label{aff23}
\and
Universit\'e C\^{o}te d'Azur, Observatoire de la C\^{o}te d'Azur, CNRS, Laboratoire Lagrange, Bd de l'Observatoire, CS 34229, 06304 Nice cedex 4, France\label{aff24}
\and
Department of Physics and Astronomy, University of the Western Cape, Bellville, Cape Town, 7535, South Africa\label{aff25}
\and
SRON Netherlands Institute for Space Research, Landleven 12, 9747 AD, Groningen, The Netherlands\label{aff26}
\and
Kapteyn Astronomical Institute, University of Groningen, PO Box 800, 9700 AV Groningen, The Netherlands\label{aff27}
\and
Tianjin Normal University, Binshuixidao 393, Tianjin 300387, China\label{aff28}
\and
Laboratoire Univers et Th\'eorie, Observatoire de Paris, Universit\'e PSL, Universit\'e Paris Cit\'e, CNRS, 92190 Meudon, France\label{aff29}
\and
Astrophysics Group, Blackett Laboratory, Imperial College London, London SW7 2AZ, UK\label{aff30}
\and
Dipartimento di Fisica, Sapienza Universit\`a di Roma, Piazzale Aldo Moro 2, 00185 Roma, Italy\label{aff31}
\and
INAF-Osservatorio Astronomico di Roma, Via Frascati 33, 00078 Monteporzio Catone, Italy\label{aff32}
\and
School of Mathematics and Physics, University of Surrey, Guildford, Surrey, GU2 7XH, UK\label{aff33}
\and
INAF-Osservatorio Astronomico di Brera, Via Brera 28, 20122 Milano, Italy\label{aff34}
\and
INFN, Sezione di Trieste, Via Valerio 2, 34127 Trieste TS, Italy\label{aff35}
\and
SISSA, International School for Advanced Studies, Via Bonomea 265, 34136 Trieste TS, Italy\label{aff36}
\and
Dipartimento di Fisica e Astronomia, Universit\`a di Bologna, Via Gobetti 93/2, 40129 Bologna, Italy\label{aff37}
\and
INFN-Sezione di Bologna, Viale Berti Pichat 6/2, 40127 Bologna, Italy\label{aff38}
\and
Dipartimento di Fisica, Universit\`a di Genova, Via Dodecaneso 33, 16146, Genova, Italy\label{aff39}
\and
INFN-Sezione di Genova, Via Dodecaneso 33, 16146, Genova, Italy\label{aff40}
\and
Department of Physics "E. Pancini", University Federico II, Via Cinthia 6, 80126, Napoli, Italy\label{aff41}
\and
INAF-Osservatorio Astronomico di Capodimonte, Via Moiariello 16, 80131 Napoli, Italy\label{aff42}
\and
INFN section of Naples, Via Cinthia 6, 80126, Napoli, Italy\label{aff43}
\and
Dipartimento di Fisica, Universit\`a degli Studi di Torino, Via P. Giuria 1, 10125 Torino, Italy\label{aff44}
\and
INFN-Sezione di Torino, Via P. Giuria 1, 10125 Torino, Italy\label{aff45}
\and
INAF-Osservatorio Astrofisico di Torino, Via Osservatorio 20, 10025 Pino Torinese (TO), Italy\label{aff46}
\and
European Space Agency/ESTEC, Keplerlaan 1, 2201 AZ Noordwijk, The Netherlands\label{aff47}
\and
Institute Lorentz, Leiden University, Niels Bohrweg 2, 2333 CA Leiden, The Netherlands\label{aff48}
\and
INAF-IASF Milano, Via Alfonso Corti 12, 20133 Milano, Italy\label{aff49}
\and
Centro de Investigaciones Energ\'eticas, Medioambientales y Tecnol\'ogicas (CIEMAT), Avenida Complutense 40, 28040 Madrid, Spain\label{aff50}
\and
Port d'Informaci\'{o} Cient\'{i}fica, Campus UAB, C. Albareda s/n, 08193 Bellaterra (Barcelona), Spain\label{aff51}
\and
Institute for Theoretical Particle Physics and Cosmology (TTK), RWTH Aachen University, 52056 Aachen, Germany\label{aff52}
\and
Institute of Cosmology and Gravitation, University of Portsmouth, Portsmouth PO1 3FX, UK\label{aff53}
\and
Institute for Astronomy, University of Hawaii, 2680 Woodlawn Drive, Honolulu, HI 96822, USA\label{aff54}
\and
Instituto de Astrof\'{\i}sica de Canarias, V\'{\i}a L\'actea, 38205 La Laguna, Tenerife, Spain\label{aff55}
\and
Institute for Astronomy, University of Edinburgh, Royal Observatory, Blackford Hill, Edinburgh EH9 3HJ, UK\label{aff56}
\and
Jodrell Bank Centre for Astrophysics, Department of Physics and Astronomy, University of Manchester, Oxford Road, Manchester M13 9PL, UK\label{aff57}
\and
European Space Agency/ESRIN, Largo Galileo Galilei 1, 00044 Frascati, Roma, Italy\label{aff58}
\and
Universit\'e Claude Bernard Lyon 1, CNRS/IN2P3, IP2I Lyon, UMR 5822, Villeurbanne, F-69100, France\label{aff59}
\and
Institut de Ci\`{e}ncies del Cosmos (ICCUB), Universitat de Barcelona (IEEC-UB), Mart\'{i} i Franqu\`{e}s 1, 08028 Barcelona, Spain\label{aff60}
\and
Instituci\'o Catalana de Recerca i Estudis Avan\c{c}ats (ICREA), Passeig de Llu\'{\i}s Companys 23, 08010 Barcelona, Spain\label{aff61}
\and
UCB Lyon 1, CNRS/IN2P3, IUF, IP2I Lyon, 4 rue Enrico Fermi, 69622 Villeurbanne, France\label{aff62}
\and
Mullard Space Science Laboratory, University College London, Holmbury St Mary, Dorking, Surrey RH5 6NT, UK\label{aff63}
\and
Departamento de F\'isica, Faculdade de Ci\^encias, Universidade de Lisboa, Edif\'icio C8, Campo Grande, PT1749-016 Lisboa, Portugal\label{aff64}
\and
Instituto de Astrof\'isica e Ci\^encias do Espa\c{c}o, Faculdade de Ci\^encias, Universidade de Lisboa, Campo Grande, 1749-016 Lisboa, Portugal\label{aff65}
\and
Aix-Marseille Universit\'e, CNRS, CNES, LAM, Marseille, France\label{aff66}
\and
INFN-Padova, Via Marzolo 8, 35131 Padova, Italy\label{aff67}
\and
Aix-Marseille Universit\'e, CNRS/IN2P3, CPPM, Marseille, France\label{aff68}
\and
Universit\'e Paris-Saclay, Universit\'e Paris Cit\'e, CEA, CNRS, AIM, 91191, Gif-sur-Yvette, France\label{aff69}
\and
Space Science Data Center, Italian Space Agency, via del Politecnico snc, 00133 Roma, Italy\label{aff70}
\and
INFN-Bologna, Via Irnerio 46, 40126 Bologna, Italy\label{aff71}
\and
School of Physics, HH Wills Physics Laboratory, University of Bristol, Tyndall Avenue, Bristol, BS8 1TL, UK\label{aff72}
\and
FRACTAL S.L.N.E., calle Tulip\'an 2, Portal 13 1A, 28231, Las Rozas de Madrid, Spain\label{aff73}
\and
INAF-Osservatorio Astronomico di Padova, Via dell'Osservatorio 5, 35122 Padova, Italy\label{aff74}
\and
Dipartimento di Fisica "Aldo Pontremoli", Universit\`a degli Studi di Milano, Via Celoria 16, 20133 Milano, Italy\label{aff75}
\and
Institute of Theoretical Astrophysics, University of Oslo, P.O. Box 1029 Blindern, 0315 Oslo, Norway\label{aff76}
\and
Jet Propulsion Laboratory, California Institute of Technology, 4800 Oak Grove Drive, Pasadena, CA, 91109, USA\label{aff77}
\and
Department of Physics, Lancaster University, Lancaster, LA1 4YB, UK\label{aff78}
\and
Felix Hormuth Engineering, Goethestr. 17, 69181 Leimen, Germany\label{aff79}
\and
Technical University of Denmark, Elektrovej 327, 2800 Kgs. Lyngby, Denmark\label{aff80}
\and
Cosmic Dawn Center (DAWN), Denmark\label{aff81}
\and
Universit\'e Paris-Saclay, CNRS/IN2P3, IJCLab, 91405 Orsay, France\label{aff82}
\and
Max-Planck-Institut f\"ur Astronomie, K\"onigstuhl 17, 69117 Heidelberg, Germany\label{aff83}
\and
NASA Goddard Space Flight Center, Greenbelt, MD 20771, USA\label{aff84}
\and
Department of Physics and Astronomy, University College London, Gower Street, London WC1E 6BT, UK\label{aff85}
\and
Department of Physics and Helsinki Institute of Physics, Gustaf H\"allstr\"omin katu 2, 00014 University of Helsinki, Finland\label{aff86}
\and
Universit\'e de Gen\`eve, D\'epartement de Physique Th\'eorique and Centre for Astroparticle Physics, 24 quai Ernest-Ansermet, CH-1211 Gen\`eve 4, Switzerland\label{aff87}
\and
Department of Physics, P.O. Box 64, 00014 University of Helsinki, Finland\label{aff88}
\and
Helsinki Institute of Physics, Gustaf H{\"a}llstr{\"o}min katu 2, University of Helsinki, Helsinki, Finland\label{aff89}
\and
NOVA optical infrared instrumentation group at ASTRON, Oude Hoogeveensedijk 4, 7991PD, Dwingeloo, The Netherlands\label{aff90}
\and
Centre de Calcul de l'IN2P3/CNRS, 21 avenue Pierre de Coubertin 69627 Villeurbanne Cedex, France\label{aff91}
\and
INFN-Sezione di Milano, Via Celoria 16, 20133 Milano, Italy\label{aff92}
\and
University of Applied Sciences and Arts of Northwestern Switzerland, School of Engineering, 5210 Windisch, Switzerland\label{aff93}
\and
Universit\"at Bonn, Argelander-Institut f\"ur Astronomie, Auf dem H\"ugel 71, 53121 Bonn, Germany\label{aff94}
\and
INFN-Sezione di Roma, Piazzale Aldo Moro, 2 - c/o Dipartimento di Fisica, Edificio G. Marconi, 00185 Roma, Italy\label{aff95}
\and
Dipartimento di Fisica e Astronomia "Augusto Righi" - Alma Mater Studiorum Universit\`a di Bologna, via Piero Gobetti 93/2, 40129 Bologna, Italy\label{aff96}
\and
Department of Physics, Institute for Computational Cosmology, Durham University, South Road, Durham, DH1 3LE, UK\label{aff97}
\and
Universit\'e Paris Cit\'e, CNRS, Astroparticule et Cosmologie, 75013 Paris, France\label{aff98}
\and
Institut d'Astrophysique de Paris, 98bis Boulevard Arago, 75014, Paris, France\label{aff99}
\and
Aurora Technology for European Space Agency (ESA), Camino bajo del Castillo, s/n, Urbanizacion Villafranca del Castillo, Villanueva de la Ca\~nada, 28692 Madrid, Spain\label{aff100}
\and
OCA, P.H.C Boulevard de l'Observatoire CS 34229, 06304 Nice Cedex 4, France\label{aff101}
\and
Institut de F\'{i}sica d'Altes Energies (IFAE), The Barcelona Institute of Science and Technology, Campus UAB, 08193 Bellaterra (Barcelona), Spain\label{aff102}
\and
DARK, Niels Bohr Institute, University of Copenhagen, Jagtvej 155, 2200 Copenhagen, Denmark\label{aff103}
\and
Centre National d'Etudes Spatiales -- Centre spatial de Toulouse, 18 avenue Edouard Belin, 31401 Toulouse Cedex 9, France\label{aff104}
\and
Institute of Space Science, Str. Atomistilor, nr. 409 M\u{a}gurele, Ilfov, 077125, Romania\label{aff105}
\and
Consejo Superior de Investigaciones Cientificas, Calle Serrano 117, 28006 Madrid, Spain\label{aff106}
\and
Universidad de La Laguna, Departamento de Astrof\'{\i}sica, 38206 La Laguna, Tenerife, Spain\label{aff107}
\and
Dipartimento di Fisica e Astronomia "G. Galilei", Universit\`a di Padova, Via Marzolo 8, 35131 Padova, Italy\label{aff108}
\and
Institut f\"ur Theoretische Physik, University of Heidelberg, Philosophenweg 16, 69120 Heidelberg, Germany\label{aff109}
\and
Universit\'e St Joseph; Faculty of Sciences, Beirut, Lebanon\label{aff110}
\and
Departamento de F\'isica, FCFM, Universidad de Chile, Blanco Encalada 2008, Santiago, Chile\label{aff111}
\and
Universit\"at Innsbruck, Institut f\"ur Astro- und Teilchenphysik, Technikerstr. 25/8, 6020 Innsbruck, Austria\label{aff112}
\and
Institut d'Estudis Espacials de Catalunya (IEEC),  Edifici RDIT, Campus UPC, 08860 Castelldefels, Barcelona, Spain\label{aff113}
\and
Satlantis, University Science Park, Sede Bld 48940, Leioa-Bilbao, Spain\label{aff114}
\and
Institute of Space Sciences (ICE, CSIC), Campus UAB, Carrer de Can Magrans, s/n, 08193 Barcelona, Spain\label{aff115}
\and
Department of Physics, Royal Holloway, University of London, TW20 0EX, UK\label{aff116}
\and
Instituto de Astrof\'isica e Ci\^encias do Espa\c{c}o, Faculdade de Ci\^encias, Universidade de Lisboa, Tapada da Ajuda, 1349-018 Lisboa, Portugal\label{aff117}
\and
Cosmic Dawn Center (DAWN)\label{aff118}
\and
Niels Bohr Institute, University of Copenhagen, Jagtvej 128, 2200 Copenhagen, Denmark\label{aff119}
\and
Universidad Polit\'ecnica de Cartagena, Departamento de Electr\'onica y Tecnolog\'ia de Computadoras,  Plaza del Hospital 1, 30202 Cartagena, Spain\label{aff120}
\and
Centre for Information Technology, University of Groningen, P.O. Box 11044, 9700 CA Groningen, The Netherlands\label{aff121}
\and
Dipartimento di Fisica, Universit\`a degli studi di Genova, and INFN-Sezione di Genova, via Dodecaneso 33, 16146, Genova, Italy\label{aff122}
\and
Infrared Processing and Analysis Center, California Institute of Technology, Pasadena, CA 91125, USA\label{aff123}
\and
INAF, Istituto di Radioastronomia, Via Piero Gobetti 101, 40129 Bologna, Italy\label{aff124}
\and
Astronomical Observatory of the Autonomous Region of the Aosta Valley (OAVdA), Loc. Lignan 39, I-11020, Nus (Aosta Valley), Italy\label{aff125}
\and
Department of Physics, Oxford University, Keble Road, Oxford OX1 3RH, UK\label{aff126}
\and
Dipartimento di Fisica e Astronomia "Augusto Righi" - Alma Mater Studiorum Universit\`a di Bologna, Viale Berti Pichat 6/2, 40127 Bologna, Italy\label{aff127}
\and
ICL, Junia, Universit\'e Catholique de Lille, LITL, 59000 Lille, France\label{aff128}
\and
ICSC - Centro Nazionale di Ricerca in High Performance Computing, Big Data e Quantum Computing, Via Magnanelli 2, Bologna, Italy\label{aff129}
\and
Instituto de F\'isica Te\'orica UAM-CSIC, Campus de Cantoblanco, 28049 Madrid, Spain\label{aff130}
\and
CERCA/ISO, Department of Physics, Case Western Reserve University, 10900 Euclid Avenue, Cleveland, OH 44106, USA\label{aff131}
\and
Technical University of Munich, TUM School of Natural Sciences, Physics Department, James-Franck-Str.~1, 85748 Garching, Germany\label{aff132}
\and
Max-Planck-Institut f\"ur Astrophysik, Karl-Schwarzschild-Str.~1, 85748 Garching, Germany\label{aff133}
\and
Departamento de F{\'\i}sica Fundamental. Universidad de Salamanca. Plaza de la Merced s/n. 37008 Salamanca, Spain\label{aff134}
\and
Dipartimento di Fisica e Scienze della Terra, Universit\`a degli Studi di Ferrara, Via Giuseppe Saragat 1, 44122 Ferrara, Italy\label{aff135}
\and
Istituto Nazionale di Fisica Nucleare, Sezione di Ferrara, Via Giuseppe Saragat 1, 44122 Ferrara, Italy\label{aff136}
\and
Center for Data-Driven Discovery, Kavli IPMU (WPI), UTIAS, The University of Tokyo, Kashiwa, Chiba 277-8583, Japan\label{aff137}
\and
Ludwig-Maximilians-University, Schellingstrasse 4, 80799 Munich, Germany\label{aff138}
\and
Max-Planck-Institut f\"ur Physik, Boltzmannstr. 8, 85748 Garching, Germany\label{aff139}
\and
Dipartimento di Fisica - Sezione di Astronomia, Universit\`a di Trieste, Via Tiepolo 11, 34131 Trieste, Italy\label{aff140}
\and
California Institute of Technology, 1200 E California Blvd, Pasadena, CA 91125, USA\label{aff141}
\and
Department of Physics \& Astronomy, University of California Irvine, Irvine CA 92697, USA\label{aff142}
\and
Department of Mathematics and Physics E. De Giorgi, University of Salento, Via per Arnesano, CP-I93, 73100, Lecce, Italy\label{aff143}
\and
INFN, Sezione di Lecce, Via per Arnesano, CP-193, 73100, Lecce, Italy\label{aff144}
\and
INAF-Sezione di Lecce, c/o Dipartimento Matematica e Fisica, Via per Arnesano, 73100, Lecce, Italy\label{aff145}
\and
Departamento F\'isica Aplicada, Universidad Polit\'ecnica de Cartagena, Campus Muralla del Mar, 30202 Cartagena, Murcia, Spain\label{aff146}
\and
Instituto de Astrof\'isica de Canarias (IAC); Departamento de Astrof\'isica, Universidad de La Laguna (ULL), 38200, La Laguna, Tenerife, Spain\label{aff147}
\and
Instituto de F\'isica de Cantabria, Edificio Juan Jord\'a, Avenida de los Castros, 39005 Santander, Spain\label{aff148}
\and
Department of Computer Science, Aalto University, PO Box 15400, Espoo, FI-00 076, Finland\label{aff149}
\and
Instituto de Astrof\'\i sica de Canarias, c/ Via Lactea s/n, La Laguna 38200, Spain. Departamento de Astrof\'\i sica de la Universidad de La Laguna, Avda. Francisco Sanchez, La Laguna, 38200, Spain\label{aff150}
\and
Caltech/IPAC, 1200 E. California Blvd., Pasadena, CA 91125, USA\label{aff151}
\and
Ruhr University Bochum, Faculty of Physics and Astronomy, Astronomical Institute (AIRUB), German Centre for Cosmological Lensing (GCCL), 44780 Bochum, Germany\label{aff152}
\and
Univ. Grenoble Alpes, CNRS, Grenoble INP, LPSC-IN2P3, 53, Avenue des Martyrs, 38000, Grenoble, France\label{aff153}
\and
Department of Physics and Astronomy, Vesilinnantie 5, 20014 University of Turku, Finland\label{aff154}
\and
Serco for European Space Agency (ESA), Camino bajo del Castillo, s/n, Urbanizacion Villafranca del Castillo, Villanueva de la Ca\~nada, 28692 Madrid, Spain\label{aff155}
\and
ARC Centre of Excellence for Dark Matter Particle Physics, Melbourne, Australia\label{aff156}
\and
Centre for Astrophysics \& Supercomputing, Swinburne University of Technology,  Hawthorn, Victoria 3122, Australia\label{aff157}
\and
DAMTP, Centre for Mathematical Sciences, Wilberforce Road, Cambridge CB3 0WA, UK\label{aff158}
\and
Kavli Institute for Cosmology Cambridge, Madingley Road, Cambridge, CB3 0HA, UK\label{aff159}
\and
IRFU, CEA, Universit\'e Paris-Saclay 91191 Gif-sur-Yvette Cedex, France\label{aff160}
\and
Oskar Klein Centre for Cosmoparticle Physics, Department of Physics, Stockholm University, Stockholm, SE-106 91, Sweden\label{aff161}
\and
INAF-Osservatorio Astrofisico di Arcetri, Largo E. Fermi 5, 50125, Firenze, Italy\label{aff162}
\and
Centro de Astrof\'{\i}sica da Universidade do Porto, Rua das Estrelas, 4150-762 Porto, Portugal\label{aff163}
\and
HE Space for European Space Agency (ESA), Camino bajo del Castillo, s/n, Urbanizacion Villafranca del Castillo, Villanueva de la Ca\~nada, 28692 Madrid, Spain\label{aff164}
\and
Department of Astrophysical Sciences, Peyton Hall, Princeton University, Princeton, NJ 08544, USA\label{aff165}
\and
Department of Astrophysics, University of Zurich, Winterthurerstrasse 190, 8057 Zurich, Switzerland\label{aff166}
\and
INAF-Osservatorio Astronomico di Brera, Via Brera 28, 20122 Milano, Italy, and INFN-Sezione di Genova, Via Dodecaneso 33, 16146, Genova, Italy\label{aff167}
\and
Theoretical astrophysics, Department of Physics and Astronomy, Uppsala University, Box 516, 751 37 Uppsala, Sweden\label{aff168}
\and
Mathematical Institute, University of Leiden, Einsteinweg 55, 2333 CA Leiden, The Netherlands\label{aff169}
\and
Leiden Observatory, Leiden University, Einsteinweg 55, 2333 CC Leiden, The Netherlands\label{aff170}
\and
Institute of Astronomy, University of Cambridge, Madingley Road, Cambridge CB3 0HA, UK\label{aff171}
\and
Department of Physics and Astronomy, University of California, Davis, CA 95616, USA\label{aff172}
\and
Center for Computational Astrophysics, Flatiron Institute, 162 5th Avenue, 10010, New York, NY, USA\label{aff173}}    

\date{}

\titlerunning{\Euclid 2D cosmic web}
\authorrunning{Euclid Collaboration: N. Malavasi et al.}

\abstract{Cosmic filaments are prominent structures of the matter distribution of the Universe. Modern detection algorithms are an efficient way to identify filaments in large-scale observational surveys of galaxies. Many of these methods were originally designed to work with simulations and/or well-sampled spectroscopic surveys. When spectroscopic redshifts are not available, the filaments of the cosmic web can be detected in projection using photometric redshifts in slices along the Line of Sight, which enable the exploration of larger cosmic volumes. However, this comes at the expense of a lower redshift precision. It is therefore crucial to assess the differences between filaments extracted from exact redshifts and from photometric redshifts for a specific survey. We apply this analysis to capture the uncertainties and biases of filament extractions introduced by using the photometric sample of the Euclid Wide Survey. The question that we address in this work is how can we compare two filament samples derived with redshifts of different precisions in the Euclid Wide Survey context. We apply the cosmic web detection algorithm DisPerSE, in the redshift range $0.1 \leq z \leq 0.5$, to the GAlaxy Evolution and Assembly (GAEA) simulated galaxy sample which reproduces several characteristics of the Euclid Wide Survey. We then explore the space of parameters that can be modified during the cosmic web reconstruction in our analysis. We develop a method to compare skeletons derived from photometric redshifts to those derived from true galaxy positions. This method expands the commonly used measure of distance between filaments to include geometrical (angles between filaments) and astrophysical considerations (galaxy mass gradients and connectivity-mass relations). We design our approach to match 2D filament samples, generated by projecting galaxies with photometric redshifts within different slices, with projected 3D filament samples reconstructured from galaxies with true positions. We assess whether this approach strengthens our ability to correctly identify filaments in very large surveys such as the Euclid Wide Survey. Additionally, we offer an overview of how the performance of filament reconstruction using photometric redshifts varies with the adjustable parameters used in the process. This paper is one of a series of three that aim to assess the quality of the cosmic web reconstruction that can be achieved with \Euclid.}

\maketitle

\section{Introduction}
\label{sec:Intro}
The cosmic web (\citealt{Gregory1978}, \citealt{Joeveer1978}, \citealt{deLapparent1986}, \citealt{Giovanelli1986}, \citealt{GellerHuchra1989}, \citealt{Shectman1996}, \citealt{BKP1996}, see also, e.g., \citealt{Feldbrugge2018}, \citealt{Feldbrugge2025}) is the complex and dynamic web-like network of matter that fills the Universe at all overdensities. Phenomenologically, it consists of galaxy clusters at the positions of the nodes of the network, connected by filaments. These are linear structures of matter, with a lower density contrast than clusters, a few Mpc to a hundred of Mpc long. Other structures of the cosmic web include walls (planar structures, see those identified, e.g., in \citealt{Gott2005}, \citealt{Iovino2016}, and \citealt{Kraljic2018}) that surround voids, large empty regions in space \citep{Colberg2008, vdWPlaten2011, Cautun2018}.

The structures of the cosmic web are composed of dark matter, gas, and galaxies. Among these different phases, the galaxy distribution offers the most immediate way to detect the cosmic web in large volumes. Compared to the gas phase (traced,e.g., by X-ray observations), it allows us to span a wider range of density contrasts and to build larger statistical samples of cosmic web elements faster with large-scale, wide-area observations. Indeed, several observational campaigns have succeeded in detecting clusters and filaments from the galaxy distribution, such as the 2 degree Field and the 6 degree Field Galaxy Redshift Surveys (2dFGRS and 6dFGRS, respectively, \citealt{Colless2001, Jones2004}), the Sloan Digital Sky Survey (SDSS, \citealt{York2000}), the Galaxy and Mass Assembly survey (GAMA, \citealt{Driver2009}), the VIMOS Public Extragalactic Redshift Survey (VIPERS, \citealt{Scodeggio2018}), the Cosmic Evolution Survey (COSMOS, \citealt{Scoville2007, Laigle2016}), and the Canada-France-Hawaii Telescope Legacy Survey (CFHTLS, \citealt{Coupon2009}).

Throughout the years, several algorithms and software have been developed to detect the cosmic web, which rely on a variety of statistical, mathematical, geometrical, and physical methods (see, e.g., \citealt{Libeskind2018} for a review). These algorithms, such as for example the Minimal Spanning Tree (MST, \citealt{Barrow1985}, \citealt{Alpaslan2014}), T-ReX \citep{Bonnaire2020}, the Subspace Constrained Mean Shift (SCMS, \citealt{Chen2015a}), BISOUS \citep{Tempel2014}, the Multiscale Morphology Filter (MMF, \citealt{AragonCalvo2007}), NEXUS+ \citep{Cautun2013}, the Discrete Persistent Structure Extractor (\disperse, \citealt{Sousbie2011a, Sousbie2011b}), SpineWeb \citep{AragonCalvo2010spineweb}, ORIGAMI \citep{Falck2012, Abel2012, Shandarin2012}, and T-web and V-web \citep{Hahn2007, ForeroRomero2009, Hoffman2012}, have been applied with success to the galaxy surveys introduced above and/or to numerical simulations of the Universe. 

Thanks to the combination of these powerful methods and ever-growing data sets, our knowledge of the cosmic web has significantly increased. Investigations range from the basic description of the cosmic web structures \citep{Colberg2005, AragonCalvo2010, Cautun2014}, their properties such as transverse thickness of filaments \citep{GalarragaEspinosa2020, GalarragaEspinosa2022, GalarragaEspinosa2023, GalarragaEspinosa2024, WeiWang2024, Tanimura2020b, Tanimura2020a, Tanimura2022, Bonjean2020, Zhang2024}, to the study of the topology and connections of the cosmic web and the impact of filaments on galaxies. They have established our view at low redshift that galaxies which are closer to the axes of filaments are more massive and less star-forming than those which are far away \citep{Alpaslan2016, Kuutma2017, Chen2017, Malavasi2017, Laigle2018, Kraljic2018, Kraljic2019, Bonjean2020, Rost2020, Salerno2020, Malavasi2022spin}. In addition, the filament environment also affects the angular momentum acquisition in galaxies, ultimately contributing to determine the direction of their spin vector (an effect which is also reflected on galaxy morphology). This is most visible in the different alignment of the major axis of disc and elliptical galaxies or in a transition mass above which galaxies have their spin perpendicular to the filaments of the cosmic web and below which the spin of galaxies is parallel to the filaments of the cosmic web (\citealt{AragonCalvo2007spin, GV2018, Codis2012, Kraljic2020b, Codis2018, Hahn2010, Malavasi2022spin, Welker2020, Trujillo2006, Paz2008, TempelLibeskind2013, Tempel2013, Chen2015b}, but see, e.g., \citealt{Zhang2013, Chen2019, Pahwa2016, GV2019, Krolewski2019, PengWang2020} for works which did not detect any transition). These examples demonstrate the growing interest in the topic of the co-evolution of galaxies and the cosmic web, demanding accurate measurements of both.

Additionally, cosmic web measurements are also used to constrain cosmological models and the history of the Universe and its content. Tracing the cosmic web in its full extent allows us to study its anisotropic properties, which, e.g., galaxy clustering alone is unable to track. An example is the measurement of the number of filaments connected to a given node or its connectivity or multiplicity \citep{Colombi2000, Pichon2010, Codis2018connectivity, Cadiou2020}. Theory and simulations show that connectivity and multiplicity can be expected to increase with node mass or environmental density \citep{AragonCalvo2010, Codis2018connectivity, Gouin2021, Kraljic2022Weave, Malavasi2023ComaSim}. Moreover node connectivity is dependent on the cosmological model \citep{Codis2013, Codis2018connectivity, Boldrini2024, Mainieri2024}. In observations, connectivity and multiplicity have been explored at cluster scales (i.e., the number of filaments connected to a given cluster, in relation to its mass, dynamical status, and galaxy population, see, e.g., \citealt{Sarron2019, DarraghFord2019, Lee2019, Malavasi2020Coma, Einasto2020A2142, Gouin2020, Castignani2022VirgoI, Castignani2022VirgoII}) and galaxy scales (i.e., the number of filaments connected to a given galaxy in relation to galaxy mass and properties, see, e.g., \citealt{Kraljic2020a, DarraghFord2019}, see also \citealt{GalarragaEspinosa2023flows} for a study on numerical simulations, \citealt{Martin2016} for the study of filaments connected to a massive proto-disc at high redshift from the gas phase or \citealt{BennettSijacki2020, Valentini2021} for examples of studies conducted on massive high-redshift haloes in numerical simulations).

The impact of the cosmic web on galaxy properties is much less known beyond $z \sim 1$, where we currently lack sufficient statistics of high-quality measurements. In order to perform these investigations, galaxy positions need to be known, including their redshifts which are essential to accurately recover the galaxy density field. Generally, spectroscopic redshifts (spec-$z$s) are the best choice to perform environmental studies beyond the nearby Universe, due to their high precision. In their absence, the more readily available photometric redshifts (photo-$z$s) may be used instead. While they are available over larger areas and for larger samples of galaxies than spec-$z$s, they also have much larger uncertainties (usually two to three orders of magnitude larger than for spec-$z$s, see, e.g., \citealt{Salvato2019} for a review)\footnote{Some works (see, e.g., \citealt{AragonCalvo2015}, \citealt{Shuntov2020}, and \citealt{Tsaprazi2023}) also explore the possibility to use spec-$z$s and photo-$z$s in combination. In particular, information on galaxy locations in the cosmic web when precisely extracted using spec-$z$s can be used to improve the estimate of galaxy photo-$z$s.}. A large uncertainty in galaxy positions along the line of sight usually translates to poor identification of the cosmic web. When only the local galaxy density field is of interest (i.e., with no intention to detect structures such as clusters or filaments), determining the effect of redshift precision can be readily investigated through numerical simulations (see, e.g., \citealt{Cooper2005, Muldrew2012, Fossati2015, Malavasi2016, Cucciati2016}).

Any analysis of the cosmic web, however, is confronted with the complexity and emergence of the system. This complexity is evident from its connectivity, lack of symmetries, wide range of densities and multiscale nature. In addition, we need to understand and characterise the differences that stem from different inputs (e.g., photo-$z$ vs spec-$z$) and different methodologies (e.g., cosmic web extraction methods). How do these variations impact the measurements, such as filament positions, orientations and connectivity, that we use in our analyses and therefore our results?  

Currently, there is no established method for comparisons of networks that stem from different extraction methods, or, indeed, for identifying the most suitable parameters for such an extraction. In recent years, several works have devised methods for comparing outputs of filament extractions. A commonly adopted method is to measure distances between filaments (or sections of filaments) in different skeleton realizations performed in the same region of space (see, e.g., \citealt{Malavasi2017, Laigle2018, Kuchner2020, Kuchner2021, Zakharova2023}). Other works have resorted to comparing distributions of filament properties such as length, node connectivity or multiplicity, node density distributions, volume and mass filling fractions, and structures associated to a given galaxy (e.g., \citealt{Libeskind2018, Malavasi2020Catalogue, Malavasi2023ComaSim, Rost2020, Zakharova2023, Cornwell2022, Cornwell2023, Bonnaire2020}) or to use other metrics (e.g., information theory, \citealt{Leclercq2016}). With the number of surveys and algorithms suitable for detecting the cosmic web rapidly increasing and the number of available catalogues of cosmic web elements correspondingly expanding, having a clear method to univocally compare sets of filaments is a strong necessity.

Motivated by the possibility of extracting filaments of the cosmic web from galaxy distributions provided by the Euclid Wide Survey, we set out to develop a robust method to compare filament sets and identify the best possible output based on \Euclid's photometric redshifts. \Euclid \citep{Laureijs2011, MellierOverview} is an European Space Agency (ESA) mission launched in 2023 with the goal of improving our understanding of the nature of dark energy and mapping the distribution of dark matter in the Universe. This will be achieved by means of two surveys, namely a Wide and a Deep component. The Euclid Wide Survey (EWS, \citealt{Scaramella2021}) is currently targeting $14\,000\:\deg^2$ of extragalactic sky aiming to provide slitless spectroscopy for tens of millions of galaxies in the redshift range $0.9 \leq z \leq 1.8$ (down to a sensitivity limit in H\,$\alpha$ of $2 \times 10^{-16}\:\mathrm{erg}\:\mathrm{s}^{-1}\:\mathrm{cm}^{-2}$). This will be complemented by photo-$z$ measurements (see, e.g., \citealt{Desprez2020}) over a much wider redshift range, obtained through a combination of photometric observations performed by \Euclid and ground data.

Although several algorithms for the extraction of the cosmic web are available, we focus in particular on the use of \disperse\ to extract the filament spines. In order to work within a controlled experiment setting we perform our analysis using simulated lightcones specifically designed to reproduce the \Euclid survey. Our approach is to detect the skeleton using simulated galaxies' true redshifts (true-$z$s, i.e., cosmological redshifts, which precisely determine the line-of-sight distance to a galaxy for a given cosmology) and their photo-$z$s. We then compare the two within a framework that allows us to provide a quantitative measurement of how similar two sets of filaments are to each other. In this way, we identify matches of photo-$z$ derived skeletons and their comparable `true' reference skeletons. Note that we do not consider skeletons extracted using spec-$z$s and that our true-$z$s do not include the contributions of galaxies' peculiar velocities or observational uncertainties on the redshift measurements. This is because we aim at comparing the performance of skeleton extraction using photo-$z$s to a reference set of `true' filaments extracted from `positional redshifts' that are not affected by redshift uncertainties. We dedicate two other papers to the performance of skeleton extractions using spec-$z$s, both in the volume of the Euclid Deep Survey (Kraljic et al., in preparation) and around selected clusters in the Euclid Deep Survey (Sarron et al., in preparation).

This paper is structured as follows: we describe the simulated data in Sect. \ref{sec:data} and the method we employ to extract the filaments of the cosmic web in Sect. \ref{sec:method}. We then set up our framework to establish skeleton similarity in Sects. \ref{sec:quantities}, \ref{sec:normdist}, \ref{sec:distance}, and \ref{sec:characterizing}. We discuss and summarize our conclusions in Sect. \ref{sec:conclusions}. We perform our analysis on the GAEA mock galaxy catalogue as well as on the Flagship mock galaxy catalogue. We show results for GAEA in the main body of the text and reserve Appendix \ref{flagship} for a brief discussion on Flagship. We thus ensure that results are consistent across two independent simulations. Note that the two simulations we exploit in this paper use slightly different cosmologies. We applied the correct cosmology for each set, as stated in Sect. \ref{sec:data}. In the following we use the convention $H_{0} = 100\,h\,\mathrm{km}\,\mathrm{s}^{-1}\,\mathrm{Mpc}^{-1}$.

\section{Simulated data}
\label{sec:data}
This work is focused on the analysis of the cosmic web extracted using photometric redshifts measured in the context of the EWS. Given the importance that EWS photometric redshifts play in our analysis, we briefly introduce them before describing the simulated data sets that we use.

The EWS is expected to detect over 1 billion galaxies for weak-lensing tomography, each of which will have photometric measurements in the $g$, $r$, $i$, $\IE$, $z$, $\YE$, $\JE$, and $\HE$ bands \citep[with $g$, $r$, $i$, and $z$ coming from LSST and UNIONS and $\IE$, $\YE$, $\JE$, $\HE$ acquired directly by \Euclid,][]{Scaramella2021, MellierOverview}. In addition, the EWS includes several \Euclid\ Calibration Fields such as AEGIS, CDFS, COSMOS, \Euclid self-calibration field (SelfCal), \Euclid ultra-deep field, GOODS-North, SXDS, and VVDS, which serve calibration purposes \citep{MellierOverview}.

The \Euclid pipeline will compute photo-$z$s for this vast dataset with two main requirements: to achieve a mean redshift scatter within $0.05(1+z)$ while keeping catastrophic outliers below $10\%$, and to determine the mean redshift in a tomographic bin with an accuracy $\sigma_{\langle z \rangle}<0.002(1+z)$ \citep{Laureijs2011, MellierOverview}. The chosen algorithm to compute photo-$z$s is the Nearest-Neighbour Photometric Redshifts (NNPZ). It is a development of the NNPZ algorithm used in the Hyper Suprime-Cam (HSC) photometric-redshift challenge \citep{Tanaka2018} that uses photometric redshift probability distribution functions (PDFzs) as labels; it has been applied to the \Euclid photometric data challenge \citep{Desprez2020}. Being a machine learning approach, it requires a reference sample for training. In the case of \Euclid, this reference sample is built using the COSMOS and \Euclid SelfCal fields, which offer photometry with a 5 to 8 times higher signal-to-noise ratio (S/N) compared to the average S/N across the EWS and additional photometric bands \citep{MellierOverview}. Photo-$z$s are computed for sources in the reference sample using Phosphoros (Paltani et al., in prep.), a Bayesian template-fitting software. While Phosphoros is capable of customising reference templates, for this particular analysis, PDFzs were generated using 33 COSMOS templates \citep{Ilbert2013}. These reference photo-$z$s serve as a training set for the NNPZ. As the training sample is built using a template-fitting algorithm, NNPZ emulates this algorithm using a machine-learning approach, which allows us to increase significantly the computational performance of the \Euclid photo-$z$ pipeline, while preserving the scientific performance. A detailed comparative analysis of various algorithms, assessing their influence on the \Euclid photo-$z$ pipeline's performance, is documented in \citet{Desprez2020}.

In order to perform our analysis we use a simulated lightcone created to reproduce the \Euclid survey: the GAlaxy Evolution and Assembly\footnote{\url{https://sites.google.com/inaf.it/gaea/home}} (GAEA) theoretical model run on the Millennium Simulation. In order to check the validity of our results we also compare with the Flagship Mock Galaxy Catalogue. We provide a brief overview of GAEA relevant to this work in the following. We introduce the Flagship lightcone in Appendix \ref{flagship}.

The GAEA ECLQ lightcone (\citealt{Zoldan2017, Fontanot2020}, see also \citealt{Hirschmann2016, DeLucia2024}) was created by applying the GAEA semianalytic model to the Millennium $N$-body simulation \citep{Springel2005}. The box of the simulation has a side length of $500 h^{-1} \mathrm{Mpc}$ filled with $2160^3$ particles (i.e., a particle mass resolution of $8.6 \times 10^8 h^{-1} M_{\sun}$)  and it was run using values for the cosmological parameters of $h = 0.73$, $\Omega_{\mathrm{m}} = 0.25$, and $\sigma_{8} = 0.9$ \citep{Sanchez2006}. Dark matter merger trees from Millennium allow us to track the evolution of the progenitors and descendants of each simulated galaxy.

The GAEA theoretical prescription describes the evolution of galaxies at different times and in different environments by modelling the exchange of mass and energy among the different baryonic components. The calibration of free parameters was performed against a selected set of observables. The GAEA ECLQ model includes a detailed treatment for non-instantaneous chemical enrichment \citep{DeLucia2014}, and a prescription for stellar feedback partly based on hydrodynamical simulations. It also includes an improved treatment of cold gas accretion onto supermassive black holes, of AGN-driven winds, and of angular momentum exchanges between different galactic components \citep{Fontanot2020, Xie2020}. The result is that the GAEA semianalytic model is able to reproduce the evolution of the galaxy stellar mass function and of the cosmic star-formation rate up to $z \lesssim 7$ \citep{Fontanot2017}, the evolution of the mass-metallicity relation and its secondary dependencies \citep{DeLucia2020, Fontanot2021}, as well as the properties of the AGN populations up to $z \sim 4$.

The lightcone itself has a diameter of $5.27 \deg$ (i.e., it covers an area on the plane of the sky of $\sim 22 \deg^2$) and it covers the redshift range $z \in [0,4]$. It was produced adopting a \citet{Chabrier2003} initial mass function and includes all galaxies with magnitude $H \leq 25$. Available galaxy parameters include positions, true redshifts, photometry in several bands (including \Euclid bands), bulge/disc decomposition measures, and central/satellite classifications. Photo-$z$s were computed for the galaxies in the lightcone with $\VIS \leq 25.0$, $\YE, \JE, \HE \leq 23.5$ (brighter than the EWS magnitude limits, Cucciati et al., private communication) using the Phosphoros code (see \citealt{Desprez2020} and references therein). These photo-$z$s take into account the photometric noise expected in the \Euclid bands complemented by the ground-based ones that will be available at the time of the Data Release 3 in the southern hemisphere \citep{Scaramella2021}. Photo-$z$s were computed using dust-attenuated magnitudes, with the same set-up as planned for the \Euclid photometry. The full photo-$z$ probability distribution function (PDFz) is saved for every galaxy in the lightcone. 

At variance with works such as \citet{JascheWandelt2012}, \citet{Shuntov2020}, or \citet{Tsaprazi2023}, there is no need for our work to implement the use of a luminosity function in our analysis. In fact, a luminosity function is implemented at the mock catalog creation level that results in a realistic galaxy distribution. We also do not implement masks reproducing the \Euclid footprint in our setup. While we do recognize that masked bright stars and the observation dithering pattern could create gaps and holes in our galaxy distribution which could impact the reconstruction of the cosmic web, taking them into account in an accurate way is a difficult task, beyond the goal of this work, and requiring an ad-hoc analysis.

\section{Method}
\label{sec:method}
We first outline the procedure for extracting filaments with \disperse\ in a general sense before discussing the application of this procedure using true-$z$s and photo-$z$s.

\subsection{Cosmic web extractor}
\label{sec:cwextractor}
To extract the filaments of the cosmic web we rely on the \disperse\ algorithm, \citep{Sousbie2011a, Sousbie2011b}. This method allows the user to obtain a topologically motivated description of the cosmic web, with particular focus on the connection between nodes and filaments. Our starting point is the distribution of galaxies, which is used to obtain the galaxy density field using the Delaunay Tessellation Field Estimator (DTFE, \citealt{SchaapvdW2000, vdWSchaap2009}).

\disperse\ then calculates the density field's spatial discrete gradient. Points where the discrete gradient is zero (called critical points) are then identified and classified based on the eigenvalues of the Hessian of the density field into maxima, minima, and two types of saddles, of which one bound to filaments -- so-called `type-2 saddles', which are local density minima within filaments -- are of interest for our analysis. Filaments are extracted as critical lines tangent to the gradient field in every point, starting at saddle and ending at maxima. 

After having identified filaments, \disperse\ applies persistent homology theory and topological simplification to the skeleton to identify structures which are likely introduced by Poisson noise in the galaxy distribution at the origin of the density field. This is achieved by pairing critical points together based on their persistence ratio (i.e., the amount of change in the topology of the density field which is necessary for a given couple to appear and disappear in the density field landscape). The persistence distribution for critical points extracted from the (simulated) data is internally compared by \disperse\ to the persistence distribution extracted from a skeleton computed on a Poisson distribution (i.e., a density field where only noise is present). Filaments are then eliminated from the skeleton if they are closer to the noise skeleton persistence distribution by a given number of standard deviations, which is set by the user. This parameter (which we call \nsig~in the following) is essentially the only parameter which can be tuned by the user in the \disperse\ cosmic web extraction\footnote{Some works (e.g., \citealt{Shivashankar2016}) argue in favour of a density-dependent persistence threshold, as the amount of spurious features remaining in the skeleton after cleaning may be higher in high-density environments. We defer to future works the investigation of this possibility, and rely on the \disperse\ implementation as is.}. 

In the \disperse\ formalism filaments consistently link maxima to type-2 saddles. Connectivity ($\kappa$) is thus defined as the number of filaments that depart a maximum and reach a saddle \citep{Codis2018connectivity}. However, while this is topologically correct, it gives rise to the problem of considering filaments as separate if they reach different saddles, even if they perfectly overlap for part of their path due to the shape of the gradient of the density field. In order to provide a description of the skeleton which is more physically driven, \disperse\ offers the possibility to insert artificial critical points (called bifurcations) at the point where the path of two or more filaments cross, eliminating all the overlapping portions of the filaments except for a single one (which will therefore be considered as a single structure and not several overlapping ones). When bifurcations are inserted, filaments then may no longer reach type-2 saddles directly, but may stop at one or more bifurcations. The connectivity of a node, minus the number of filaments that stop at a bifurcation when they are inserted is a measurement of the node multiplicity ($\mu$, \citealt{Codis2018connectivity}).

\subsection{Filament extraction with photo-$z$s}
In the case of photo-$z$s, their uncertainties prevent us from extracting the cosmic web in 3D. We therefore decided to run \disperse\ in 2D on the plane of the sky, as done previously in similar situations for dealing with photo-$z$ uncertainties (see, e.g., \citealt{Laigle2018, Sarron2019, DarraghFord2019}). In order to do this, we have to divide galaxies into redshift slices based on their photo-$z$. We approach this problem by first setting the slice thickness.

\begin{figure}
\centering
\includegraphics[width = \columnwidth, trim = 0cm 0cm 1cm 1cm, clip=true]{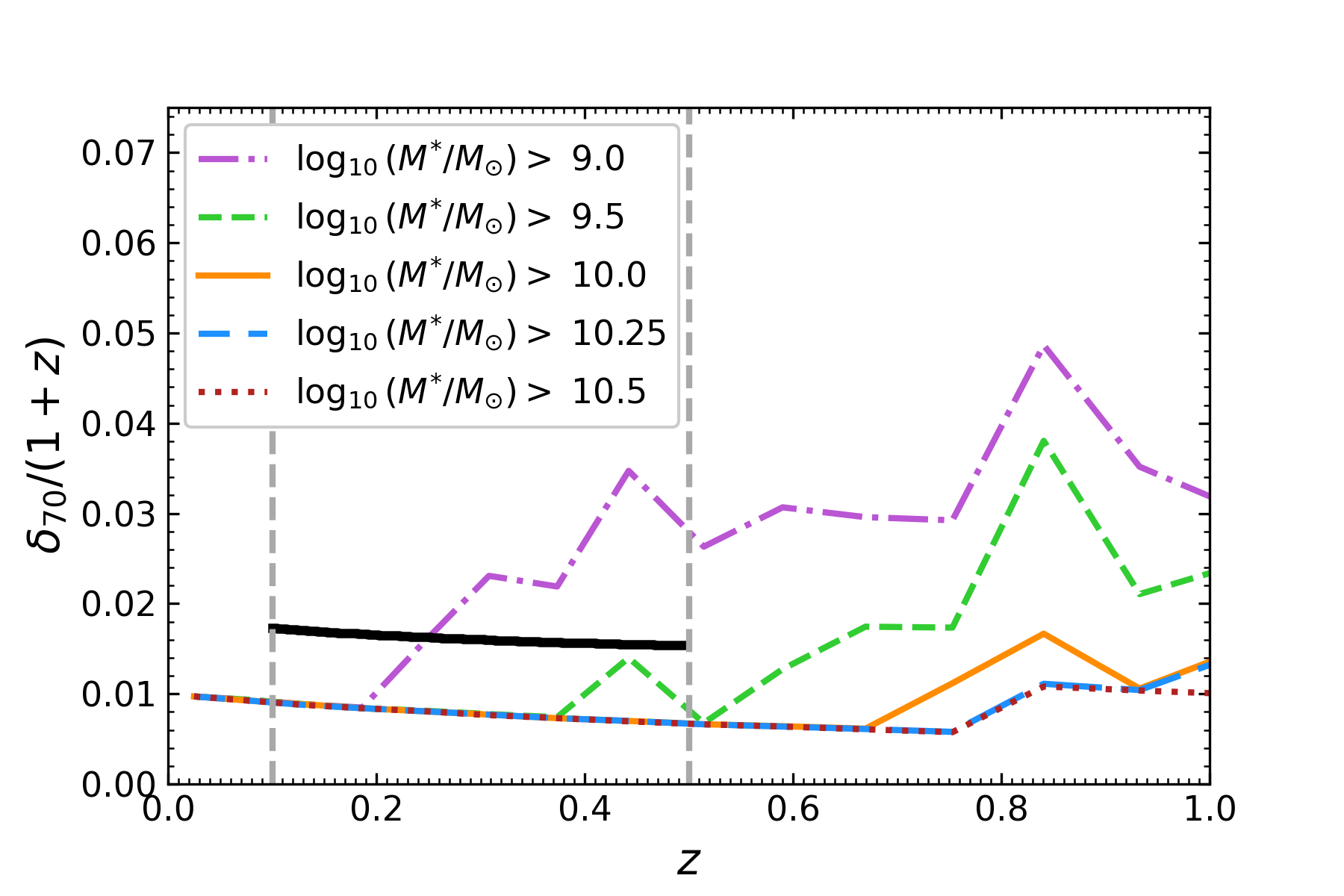}
\caption{Photo-$z$ $\delta_{70}$ interval which provides an estimate of the  $\pm 1\sigma_{z}/(1+z)$ uncertainty interval as a function of redshift. The photo-$z$ uncertainty corresponds to the $\delta_{70}$ interval encompassing 70\% of the PDFz divided by $(1+z)$, for galaxies above different stellar mass cuts: $\logten(M^{\ast}/M_{\sun}) > 9$ (purple dot-dashed line), $\logten(M^{\ast}/M_{\sun}) > 9.5$ (green short-dashed line), $\logten(M^{\ast}/M_{\sun}) > 10$ (orange solid line), $\logten(M^{\ast}/M_{\sun}) > 10.25$ (blue long-dashed line), and $\logten(M^{\ast}/M_{\sun}) > 10.5$ (red dotted line). The thick black line shows the width in redshift units of a 75 comoving Mpc thick slice as a function of redshift. The dashed vertical lines show the redshift interval considered in this work. We note that we use the $\delta_{70}/(1+z)$ instead of the $\pm 1\sigma_{z}/(1+z)$ interval to take into account the possibility of an asymmetrical photo-$z$ uncertainty due to the shape of the PDFz.}
\label{redshift_vs_sigma}
\end{figure}

Figure \ref{redshift_vs_sigma} shows the photo-$z$ uncertainty as a function of redshift for galaxies fulfilling different stellar mass cuts. It is established that photo-$z$ accuracy is related to galaxy stellar mass, as more massive galaxies tend to have a spectral energy distribution (SED) with more distinctive features. Photo-$z$ uncertainty in the case of Fig. \ref{redshift_vs_sigma} was computed by performing 100 random extractions from the photo-$z$ PDFz for each galaxy, with each extraction leading to a different redshift measurement for each galaxy. For each extraction, galaxies were then divided in bins of size $\Delta z = 0.05(1+z)$. Within each bin, we measure the interval encompassing 70\% of each PDFz, and define it as $\delta_{70}$. The median $\delta_{70}$ for the 100 PDFz are shown in Fig. \ref{redshift_vs_sigma}. We chose the $\delta_{70}$ uncertainty interval to obtain an estimate of the comoving distance corresponding to the $\pm \sigma_{z}/(1+z)$ uncertainty interval.

In this same figure we report also the width of a 75 comoving Mpc thick redshift slice as a function of redshift (thick black line in the figure). This figure shows that within the redshift interval that we considered for our work ($0.1 \leq z \leq 0.5$), this slice thickness is larger than the $\delta_{70}$ interval of photo-$z$s for galaxies more massive than $10^{10} M_{\sun}$ and it is comparable to this quantity for galaxies more massive than $10^{9.5} M_{\sun}$ at the upper edge of our redshift interval ($0.1 \leq z \leq 0.5$). We therefore consider this slice to be large enough so that most galaxies with their photo-$z$ located close to the centre of the slice are not scattered in or out of the slice itself by their uncertainty. We thus choose to use a fixed slice thickness of 75 comoving Mpc to compute our 2D skeletons. This results in 39 redshift slices from $0.1 \leq z \leq 0.5$ with mean width of 0.02 in redshift. Our slices are designed to overlap, so that the lower edge of a slice is located at the same redshift of the centre of the previous one. We choose a very conservative redshift range for our analysis to ensure that our choice of slice thickness is larger than the photo-$z$ uncertainty for our selected galaxy population. We could in principle extend our analysis to redshift $z \sim 0.7$ and our choice of 75 Mpc for the slice thickness would still be larger than the photo-$z$ uncertainty for galaxies with $M^{\ast} \geq 10^{10} M_{\sun}$. However, in this case, galaxies at the lower edge of our selected mass range (i.e., those with $10^{9.75} \leq M^{\ast}/M_{\sun} \leq 10^{10.25}$) would have a photo-$z$ uncertainty larger than our adopted redshift slice thickness in the redshift range $0.5 \leq z \leq 0.7$ (not shown in Fig. \ref{redshift_vs_sigma}).

We divide galaxies into redshift slices using the full information of the PDFz and not just their best redshift estimate. Galaxies are attributed to a redshift slice if the integral of their PDFz within the slice is greater than a threshold (which we refer to as \pt). We try several values for the parameter \pt, namely 0.2, 0.3, 0.4, 0.5, and 0.6. This corresponds to galaxies that have a probability of being within the slice ranging from 20\% to 60\%. Given these thresholds, the fraction of galaxies that belong to no slice and are therefore not used in the skeleton extraction is at most 5\% and in the majority of cases below 1\%.

Given that the photo-$z$ uncertainty depends on galaxy stellar mass (photo-$z$ are better estimated for galaxies of larger mass) we also try to impose several mass selections to the sample. In particular we select all galaxies with $M^{\ast}$ greater than a certain threshold, which we refer to as \mlim\ (with both in units of $M_{\sun}$). As with \pt, we try several values for $\logten($\mlim$/M_{\sun})$, namely 10.0, 10.25, and 10.5.

We then extract the skeleton with \disperse\ using all the galaxies with $M^{\ast} >\,$\mlim\ within a given redshift slice according to a choice of the parameter \pt.\footnote{This assumes that all galaxies within a slice are located on a plane at the redshift of the slice centre. This approximation is incorrect, as for areas as large as the final extent of the EWS galaxies should rather be projected on a plane from the spherical shell they occupy. We take this into account when projecting 3D skeletons (see next section). Comparing the performance of \disperse\ on a sphere and a plane is outside the goal of this work.} We do not apply a smoothing of the density field prior to the skeleton extraction (so as not to introduce another variable in our setup) but we do apply a smoothing of the skeleton after its extraction. This is achieved by averaging the positions of the extremes of each segment of each filament with the positions of the extremes of the neighbouring segments. The procedure is iterated twice and the positions of the critical points (i.e., the extremes of the filaments) are not modified. This minor smoothing of the filaments after their extraction is introduced to slightly reduce the local variations in the filament shapes, while keeping the global shape of the filaments unmodified. It is necessary for the purpose of filament alignment measurement. As this is a minor modification of the filament shape and does not affect the skeleton global properties, we do not feel the need to perform a full exploration of the effects of the chosen smoothing value. We perform the skeleton extraction using a variety of values for the persistence threshold which can be tuned in \disperse\ (we referred to this parameter in a general way as \nsig\ in Sect. \ref{sec:cwextractor}; we refer to it as \nsigtwo\ to specify the case of the 2D skeleton extraction and \nsigthree\ to specify the case of the 3D skeleton extraction). In particular we try values of \nsigtwo~of 2.0, 2.5, 3.0, 3.5, i.e., we select as not spurious those pairs of critical points whose persistence value is more different from the persistence distribution of a noise skeleton than 2--3.5 times its standard deviation. The choice of these values ensures that between less than 5\% and less than 0.1\% of the filaments in the considered skeleton are spurious. In the end, we test 4 values of \nsigtwo, 5 values of \pt, and 3 values of \mlim. Given that we perform a skeleton extraction for each parameter combination, this results in a total of $4 \times 5 \times 3 = 60$ skeleton extractions in each redshift slice in the 2D case. We provide a summary of the parameters we use for the 2D skeleton extraction in Table \ref{paramsummary}.

\begin{table}
\caption{Summary of the parameters used in the 2D and 3D filament extractions. In particular, \pt\ is the probability threshold used to determine if a galaxy belongs or not to a redshift slice, $\logten($\mlim$/M_{\sun})$ is the limiting mass of the sample, \nsigtwo\ and \nsigthree\ are the persistence thresholds used for the 2D and 3D-projected skeleton extractions.}
\label{paramsummary}
\centering
\begin{tabular}{c c}
Parameter & Values \\ 
\hline\hline
\multicolumn{2}{c}{2D skeleton} \\
\hline
\pt & 0.2, 0.3, 0.4, 0.5, 0.6 \\
$\logten($\mlim$/M_{\sun})$ & 10.0, 10.25, 10.5 \\
\nsigtwo & 2.0, 2.5, 3.0, 3.5 \\
\hline
\multicolumn{2}{c}{3D-projected skeleton} \\
\hline
\nsigthree & 5.5, 6.0, 7.0, 7.5, 8.0 \\
\hline
\end{tabular}
\end{table}

\subsection{Filament extraction with true-$z$s}
\label{sec:fil_extraction}
In the case of true-$z$s, we extract the \disperse~skeleton in 3D, using all galaxies with $M^{\ast} \geq 10^{9} M_{\sun}$. In this case we vary only one parameter: the persistence threshold choice (we refer to the choice of \nsig\ for the 3D skeleton as \nsigthree). After extraction, the skeleton is divided in the same redshift slices used for the 2D skeleton extraction. Only portions of filaments that fall within each redshift slice are considered, and artificial critical points are inserted at the extremities of these filaments to keep consistency with the \disperse~formalism (which prescribes that critical points are always found at the extremities of filaments, this also ensures that we would be able to identify filaments which have been partly cut due to falling outside of the redshift slice). Filaments are projected on a plane located at the centre of each redshift slice to allow a fair comparison with the 2D skeleton. We refer to the skeletons extracted with true-$z$s, divided and projected within redshift slices as 3D-projected skeletons. The slicing and projection of 3D filaments in 2D slices is necessary to ease the comparison between filament samples.

We note that only type-2 saddles (local density minima bound to filaments) and filaments themselves are considered for projection. Type-1 saddles (i.e., local density minima bound to walls) and walls are not. This ensures that in our reference 3D-projected skeleton there are no spurious filaments caused by misclassified walls due to projection effects\footnote{3D walls that when projected are misclassified as filaments could still be present in the 2D skeleton. Based on the strength of this contamination and together with the effect of photo-z uncertainty, this is a factor that causes a 2D skeleton to be more or less similar to a 3D-projected reference one.}.

We vary the persistence threshold \nsigthree~among 6 values: 5.5, 6.0, 6.5, 7.0, 7.5, 8.0. This results in 6 different skeleton extractions in each redshift slice for the 3D-projected skeletons. We provide a summary of the values we use for this parameter in Table \ref{paramsummary}. In the end, our work compares 60 2D skeletons with 6 3D-projected skeletons for a total of 360 individual skeleton comparisons within each redshift slice.

\section{Measuring skeleton properties}
\label{sec:quantities}
To fulfil our goal of devising an algorithm to determine similarity between skeletons extracted with \disperse, we first qualitatively inspect the skeleton properties that we aim to compare and match. We then design our algorithm to quantitatively pick up differences in the distribution of these properties and translate them into a quality label for the match between filament sets.

\begin{figure*}
\centering
\includegraphics[width=\textwidth, trim=4.5cm 2.5cm 4.5cm 3cm, clip=true]{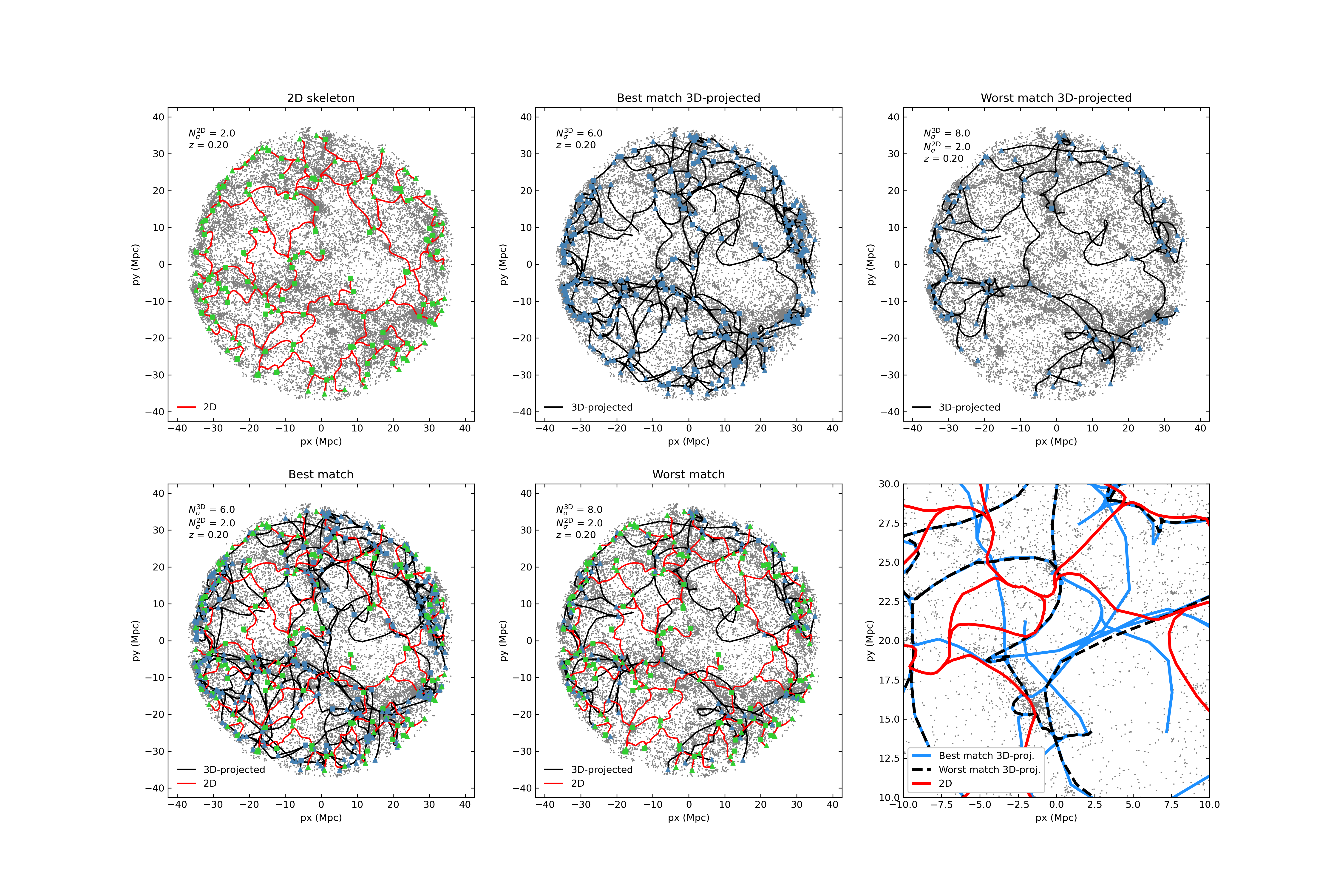}\\
\caption{Visual comparison of 2D and 3D-projected skeletons in a relevant redshift slice of width 75 Mpc (low-redshift slice, $z = 0.20$). From left to right, the panels in the top row show a 2D skeleton of a given \nsigtwo\ (chosen based on visual clarity of the filaments as an example), its corresponding best match 3D-projected skeleton (top center) and its corresponding worst match one (i.e., the skeleton couples that are the most and least similar, according to our algorithm). Bottom row shows a superposition of the same 2D skeleton as in the top leftmost panel with its best match 3D-projected skeleton (bottom left) and its worst match 3D-projected skeleton (bottom center). Bottom right panel shows a zoom in onto a random region to better detail the differences and similarities between the skeletons, with all skeletons overplot. In each panel but the bottom right, red lines and green points refer to the filaments and critical points of the 2D skeleton, black lines and blue points to the filaments and critical points of the 3D-projected skeletons. Only maxima (squares) and bifurcations (triangles) are shown for clarity. In the bottom right panel,3 red lines refer to filaments of the 2D skeleton, light blue solid lines to the filaments of the best match 3D-projected skeleton, and black dashed lines to the filaments of the worst match 3D-projected skeleton. No critical points are shown for clarity. Grey points represent the galaxy distribution within the redshift slice, galaxies are attributed to the slice according to their true-$z$. The chosen values of \nsigtwo\ and \nsigthree\ are reported in the panels. Only the case of \pt$\,= 0.4$ and $\logten($\mlim$/M_{\sun}) = 10.25$ is shown.}
\label{skelvinspect_lowz}
\end{figure*}

\begin{figure*}
\centering
\includegraphics[width=\textwidth, trim=4.5cm 2.5cm 4.5cm 3cm, clip=true]{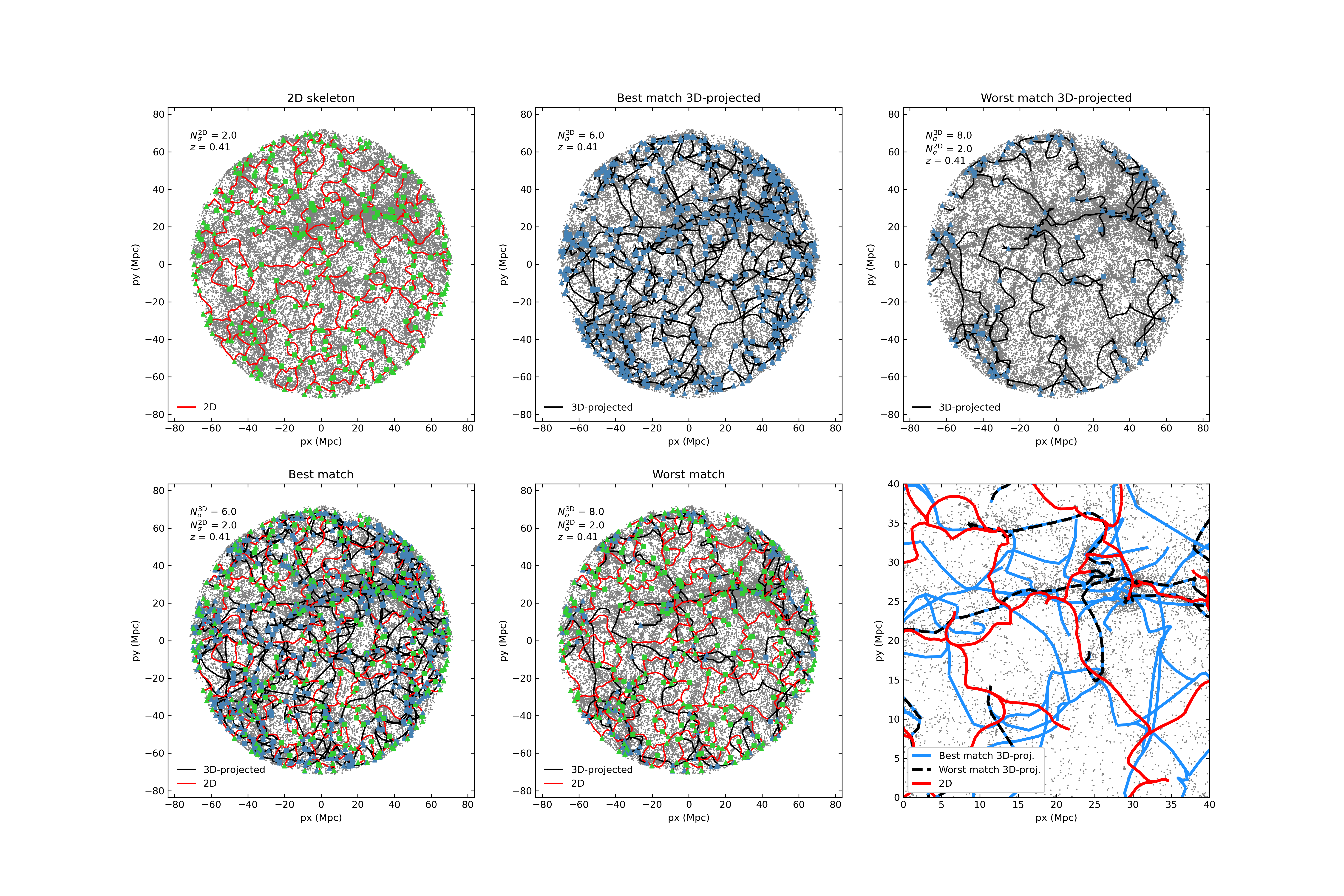}\\
\caption{Same as figure \ref{skelvinspect_lowz} but for a high-redshift slice ($z = 0.41$).}
\label{skelvinspect_highz}
\end{figure*}

Figures \ref{skelvinspect_lowz} and \ref{skelvinspect_highz} provide an example of the direct visual comparison between the 2D and the 3D-projected skeletons. These figures show filaments and critical points (maxima and bifurcations only, for clarity) belonging to a 2D and a 3D-projected skeleton, in the case of the best match and the worst match between the two, in a low-redshift ($z = 0.2$, Fig. \ref{skelvinspect_lowz}) and a high-redshift ($z = 0.41$, Fig. \ref{skelvinspect_highz}) slice, superimposed to the galaxy distribution in the slice (according to the galaxies' true-$z$s, no limit in mass is applied to the galaxies). We note that, here and in the following, whenever we show examples of the best and the worst match between filament sets, the quality of the agreement between skeletons has been set by our algorithm and has been determined as the couple of skeletons that are the most and the least similar (according to the criterion established in Sect. \ref{ddef}).

Figures \ref{skelvinspect_lowz} and \ref{skelvinspect_highz} show that (as expected) the 3D-projected skeleton better follows the galaxy distribution, which is plotted here using true-$z$s. However, even in the case of the 2D skeleton, a subset of filaments follows the galaxy distribution, especially in proximity of prominent nodes.

In the case of the best match between 2D and 3D-projected skeletons at both low and high redshift, several filaments overlap along part of their path, sharing the same orientation and position (this is particularly visible in the bottom right panels in each figure, which shows a zoom onto specific regions, with both the best and worst match 3D-projected skeletons overplot onto the chosen 2D one). In  the case of the worst match, fewer filaments share the same position in both projections, and the sets appear less similar upon visual inspection. A key factor for a good match is the level of detail and the number of fine filaments reconstructed (i.e., a combination of filament length, number of filaments in the skeleton, amount of filaments branching out of a given one). Since we used a low-persistence 2D skeleton with many filaments, the best matching skeleton is expected to have a similar number of small branches.

In order to compare skeletons, we measure and contrast several skeleton properties. In particular, our algorithm is based on the comparison of \emph{geometric properties} of the filaments (i.e., those that focus on differences in the shape of the skeleton), while we use \emph{astrophysical properties} (i.e., those that focus more on the type of astrophysical information that can be extracted from the skeleton) to further characterise the performance of our method (see Sect. \ref{sec:characterizing}). We next describe \emph{geometric} skeleton properties.

\subsection{Distance between skeleton segments}
For each segment of each filament of the 3D-projected skeleton we compute the distance to the closest segment of the closest filament of the 2D skeleton. We use the midpoints of the segments for this. We then repeat the computation starting from the other skeleton. The output of this measurement is a set of distributions of distances in Mpc, from each segment of the 3D-projected skeleton to the closest segment of the 2D one (3D $\rightarrow$ 2D) and viceversa (2D $\rightarrow$ 3D). In an ideal case of two perfect matching skeletons these distributions would be Dirac delta functions centred on zero. If the distributions do not overlap, they do not peak at the same value. This indicates that the two skeletons differ in detail, such as the number of small filaments branching from the main ones. In this case, the nearest segment in the other skeleton could be several Mpc away, and this asymmetry might not hold if the skeletons are reversed. If the distributions peak at a non-zero value or have a wide shape, it means the filaments do not overlap and are located at different positions in space, highlighting that the distance between filaments is on average above a certain threshold.

\begin{figure*}
\centering
\includegraphics[width = \textwidth, trim = 1.5cm 1cm 1.5cm 1cm, clip=true]{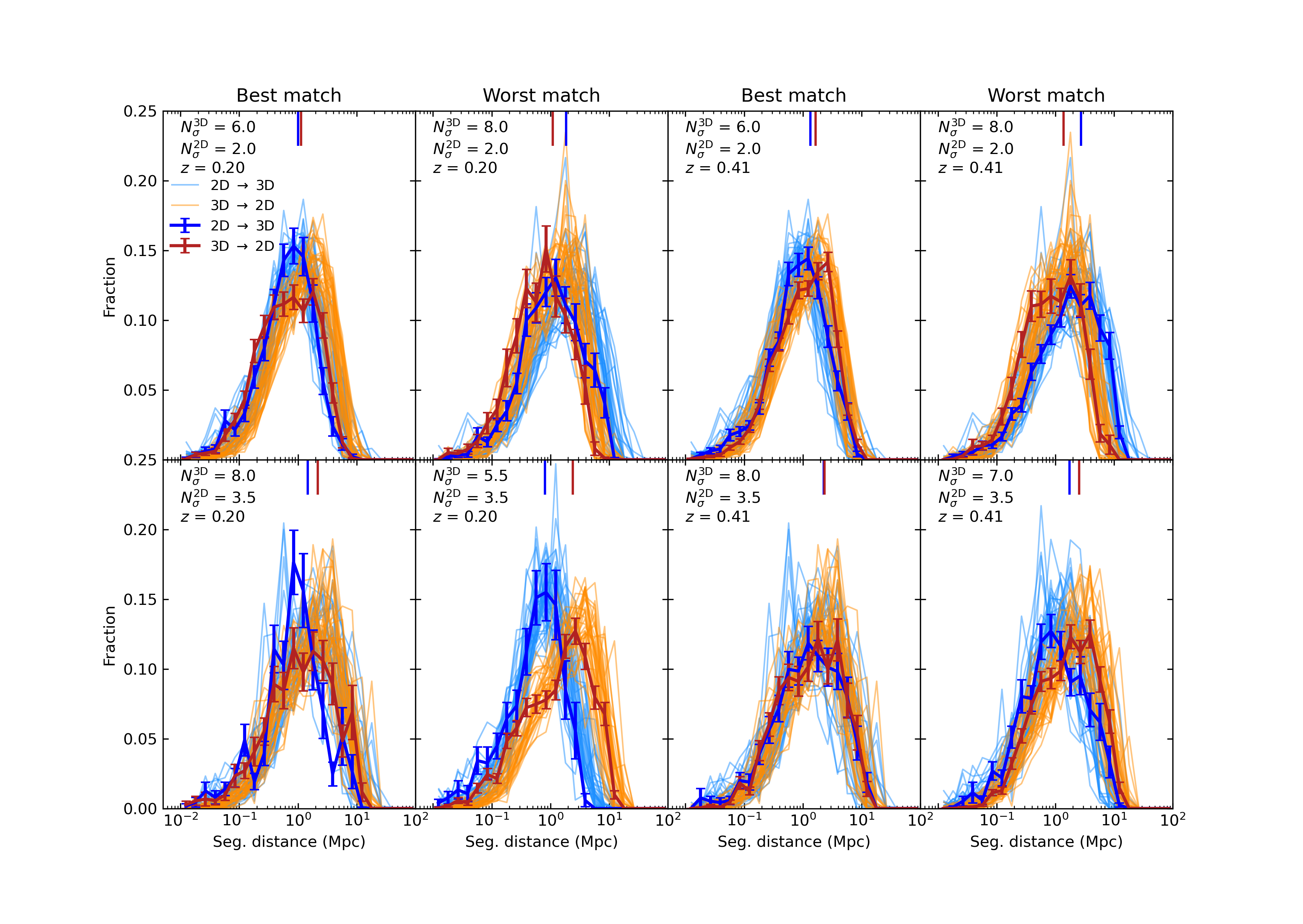}
\caption{Distributions of distances between skeleton segments. This panel refers to the case of galaxy mass limit $\logten($\mlim$/M_{\sun}) = 10.25$ and \pt$\,= 0.4$. The top (bottom) row of panels refers to the case \nsigtwo$ = 2.0$ (\nsigtwo$ = 3.5$). Columns represent either values of \nsigthree\ which correspond to a good match between the skeletons or to a bad match. The four panels on the left (right) correspond to $z = 0.2$ ($z = 0.41$). In each panel, light orange (light blue) lines refer to the case 3D $\rightarrow$ 2D (2D $\rightarrow$ 3D) at all redshifts. The lines corresponding to the relevant redshift slice shown in each panel are highlighted in dark red and dark blue. Short lines at the top of each panel represent the expectation values of the distribution of distances between skeleton segments in the slice (2D $\rightarrow$ 3D: dark blue, 3D $\rightarrow$ 2D: dark red).}
\label{distdist}
\end{figure*}

Figure \ref{distdist} shows distributions of segment distances for the best and the worst match between 2D and 3D-projected skeletons. They are derived as the fraction of segments of one skeleton with the closest segment of the other skeleton in a given distance bin divided by the total number of segments of the first skeleton, i.e., $f(\Delta d) = N_{\Delta d}/N_{\mathrm{tot}}$, where $\Delta d$ is a distance bin, $N_{\Delta d}$ the number of segments in that bin, and $N_{\mathrm{tot}}$ the total number of segments in the skeleton. We show extreme cases of a high-persistence and a low-persistence 2D skeleton, and of a high-redshift and a low-redshift slice for a fixed value of \pt\ and \mlim\ (in the middle of the range of these parameters that we explore). While the distributions for the relevant redshift slices are highlighted with thick lines, we also report the individual distributions for all the redshift slices in the background to provide a sense of the general trend. 

In all comparisons, the distributions have a rather broad shape, which indicates the presence of differences between the 3D-projected and the 2D skeletons. The peak of the distributions, even in the best match case, is not located at zero, meaning that the spatial agreement of the filament paths is not perfect. However, the peak is located between 1 Mpc and 2 Mpc for almost all the parameter combinations that we explore (including other persistence thresholds, redshifts, and values of \pt\ and \mlim\ not shown here). This indicates the typical size of the differences in the filament positions between the 2D and 3D-projected skeleton. The position of the peaks shifts farther away from zero in the worst match case between skeletons.

In the case of the best match between the skeletons, the two sets of distributions (3D $\rightarrow$ 2D and 2D $\rightarrow$ 3D) peak at roughly the same value, while in the case of the worst match between the two the peaks are shifted. As is visible from visual inspection (Figs. \ref{skelvinspect_lowz} and \ref{skelvinspect_highz}) the position of the peaks at the same location is an indication of the fact that the number of filament details are the same on average between the 3D-projected and the 2D skeleton, while the position of the individual filaments and segments may differ (which is indicated by the broad shapes of the distributions and the fact that the peaks are not exactly at zero, see similar discussions in \citealt{Malavasi2017}, \citealt{Laigle2018}, \citealt{Kuchner2020}, and \citealt{Kuchner2021}).

\subsection{Angle between skeleton segments} 
In the same way in which we measure distances between the skeleton segments, we can measure the angles between them. For each segment in the 3D-projected skeleton we consider the angle between it and the closest segment in the 2D one. The angle can range from 0 to 90 degrees. If it is close to 0 degrees it means that the two segments are aligned, while if it is close to 90 degrees, it means the two segments are perpendicular. We then switch skeletons and repeat the measurement. As before the result is a distribution of angle values in degrees, between each segment of the 3D-projected skeleton and the closest segment of the 2D one (3D $\rightarrow$ 2D) and vice versa (2D $\rightarrow$ 3D).

\begin{figure*}
\centering
\includegraphics[width = \textwidth, trim = 1.5cm 1cm 1.5cm 1cm, clip=true]{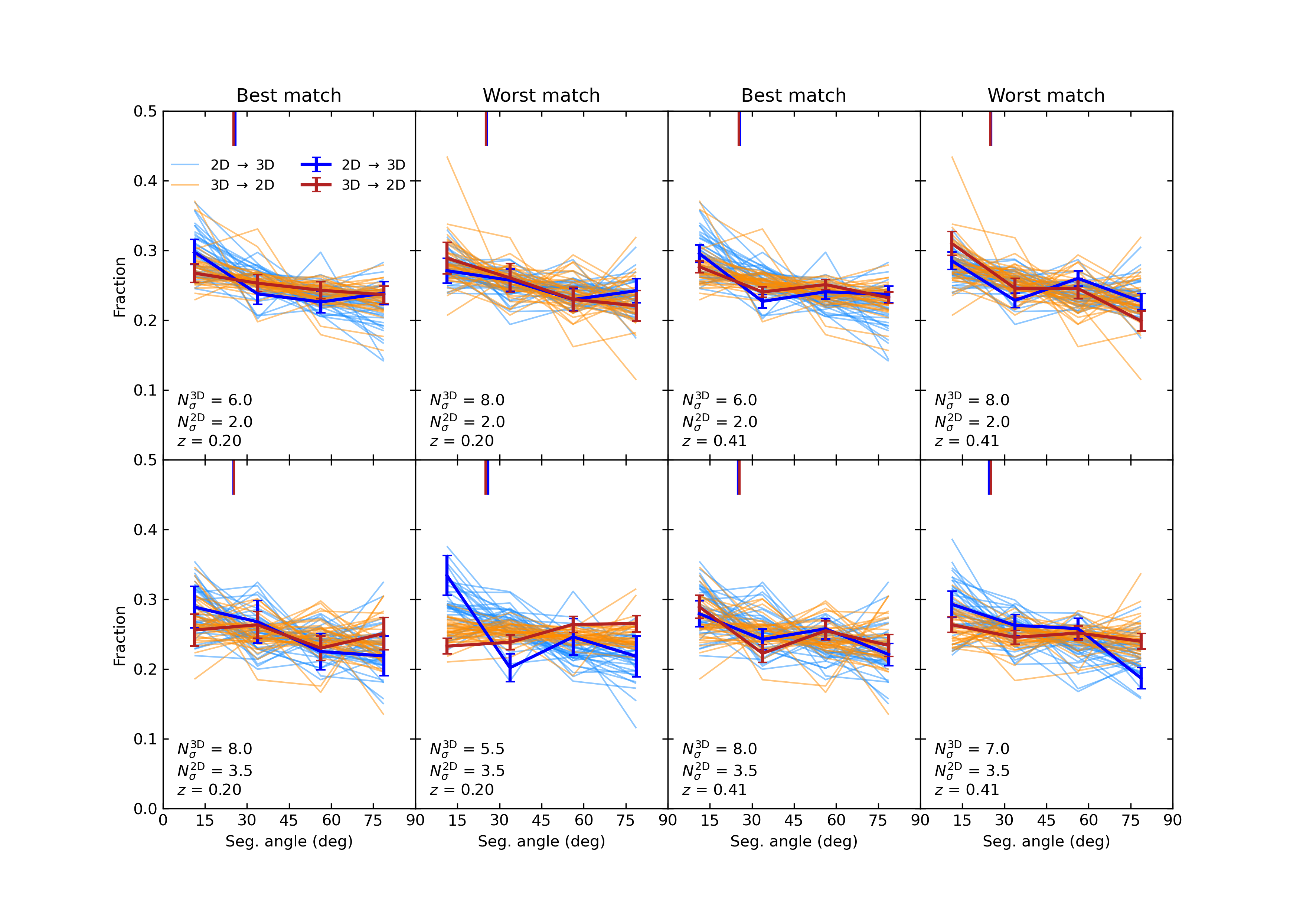}
\caption{Distributions of angles between skeleton segments. This panel refers to the case of galaxy mass limit $\logten($\mlim$/M_{\sun}) = 10.25$ and \pt$\,= 0.4$. The top (bottom) row of panels refers to the case \nsigtwo$ = 2.0$ (\nsigtwo$ = 3.5$). Columns represent either values of \nsigthree\ which correspond to a good match between the skeletons or to a bad match. The four panels on the left (right) correspond to $z = 0.2$ ($z = 0.41$). In each panel, light orange (light blue) lines refer to the case 3D $\rightarrow$ 2D (2D $\rightarrow$ 3D) at all redshifts. The lines corresponding to the relevant redshift slice shown in each panel are highlighted in dark red and dark blue. Short lines at the top of each panel represent the standard deviations of the distribution of angles between skeleton segments in the slice (2D $\rightarrow$ 3D: dark blue, 3D $\rightarrow$ 2D: dark red).}
\label{angdist}
\end{figure*}

The distributions of angles between skeleton segments are shown in Fig. \ref{angdist}. They are the fraction of segments of one skeleton with an angle from the closest segment in the other skeleton in a given angle bin, divided by the toal number of segments in the first skeleton. As for the distance case they are defined as $f(\Delta a) = N_{\Delta a}/N_{\mathrm{tot}}$ where $\Delta a$ is an angle bin, $N_{\Delta a}$ the number of segments in that bin, and $N_{\mathrm{tot}}$ the total number of segments in the skeleton. As in the case of distances between filament segments, we report only the case of $\logten($\mlim$/M_{\sun}) = 10.25$ and \pt$\,= 0.4$, for the best and worst match between skeletons, in a high- and a low-redshift slice, for a high- and a low-persistence 2D skeleton. We highlight the distributions of angles in the relevant redshift slices, while we report in the background all the distributions for all the slices to provide a sense of the general trend. These distributions have a long tail, extending all the way to 90 degrees, but also show an enhancement close to zero. This indicates that a fraction of filaments are misaligned and that overall, although each segment of the 3D-projected (2D) skeleton has a segment of a filament of the 2D (3D-projected) skeleton nearby, their direction may differ. Overall, this indicates a good global match between skeletons, although the individual shapes of the filaments differ. In addition, the difference in the angle distributions between the best and worst match cases are small. We attribute this effect to the fact that angles between segments probe differences in the local, small scale shape of the filaments. Although filaments can overlap for portions of their path, the local variations in the shape of the filaments introduced in the 2D skeleton generally do not match those present in the 3D-projected one.

\section{Residual distributions}
\label{sec:normdist}
It is challenging to consolidate the distributions discussed in the previous section as possible indicators for skeleton accuracy. They come in different shapes, normalizations and span wide absolute ranges. Combining them into a single definition of `similarity' between skeletons is not straightforward. For this reason we re-normalize the distributions of distances and angles between filament segments in the following way.

\begin{figure*}
\centering
\includegraphics[width = \textwidth, trim = 1.5cm 1cm 1.5cm 1cm, clip=true]{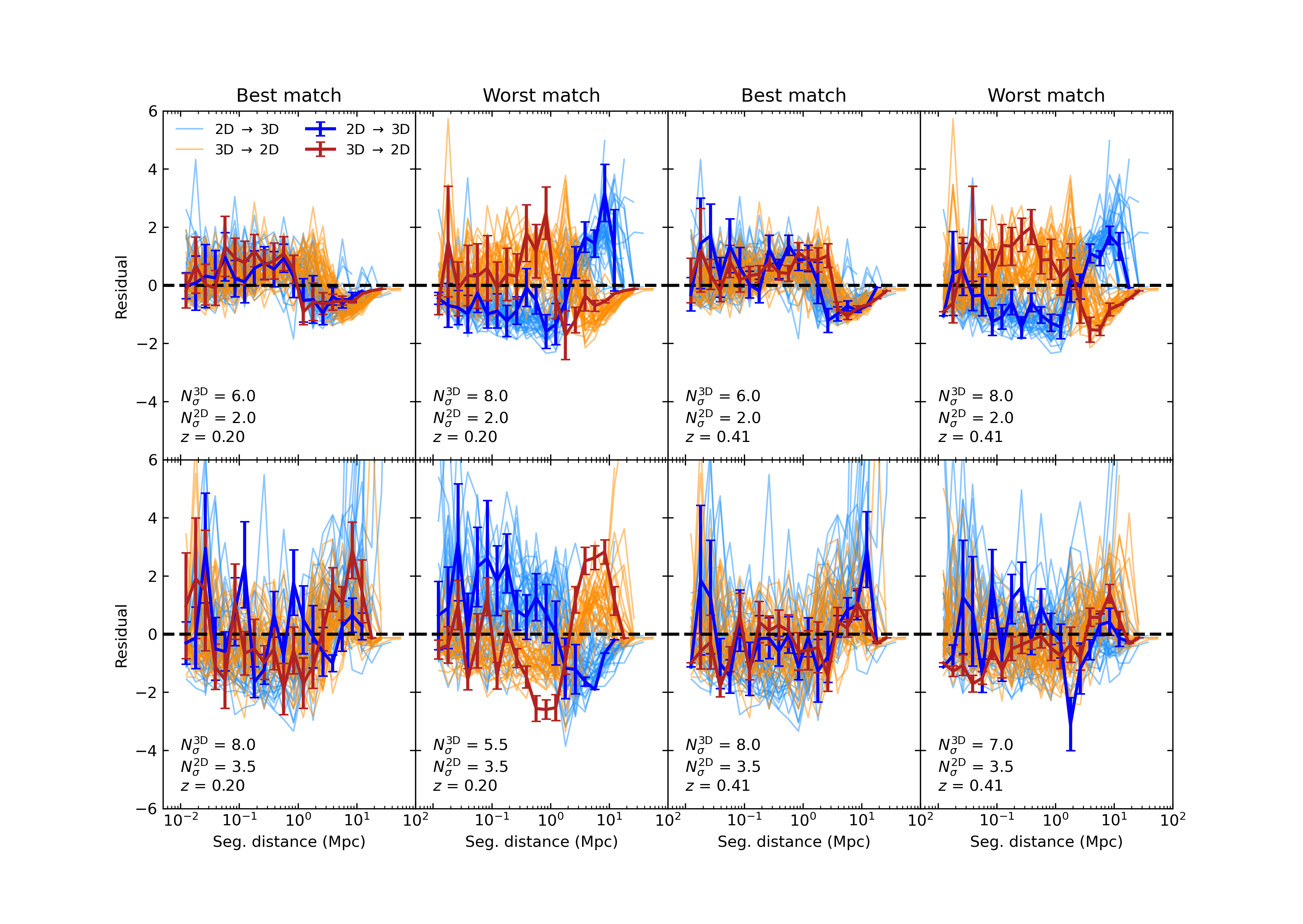}
\caption{Same as Fig. \ref{distdist}, but showing the residuals with respect to the average of the distributions of distances between skeleton segments normalized by the standard deviation, with averages and standard deviations computed across all possible choices of the parameters \mlim, \pt, \nsigtwo\ (for the 2D skeleton), and \nsigthree\ for the 3D-projected one.}
\label{distdist_normalized}
\end{figure*}

\begin{figure*}
\centering
\includegraphics[width = \textwidth, trim = 1.5cm 1cm 1.5cm 1cm, clip=true]{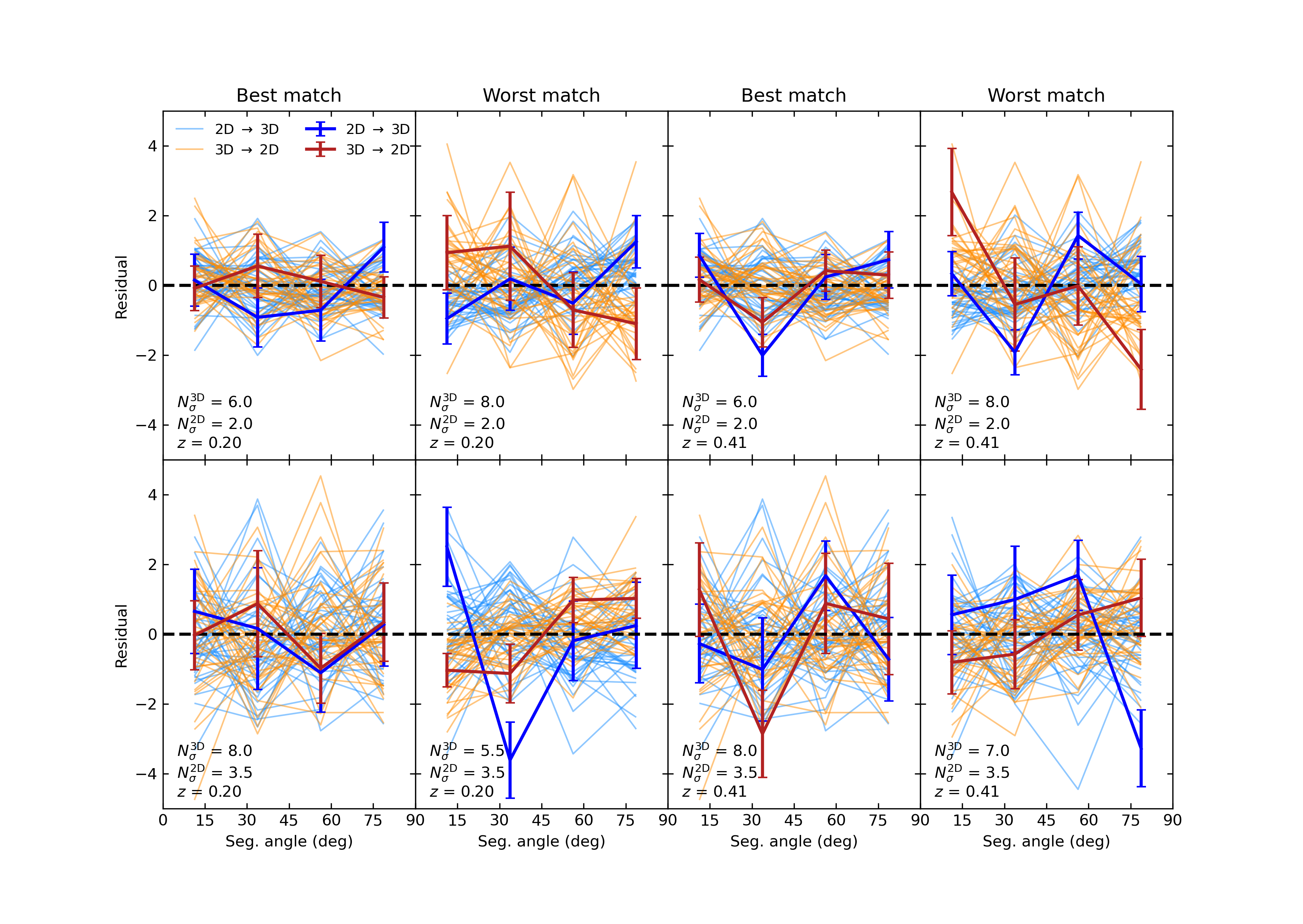}
\caption{Same as Fig. \ref{angdist}, but showing the residuals with respect to the average of the distributions of angles between skeleton segments normalized by the standard deviation, with averages and standard deviations computed across all possible choices of the parameters \mlim, \pt, \nsigtwo\ (for the 2D skeleton), and \nsigthree\ for the 3D-projected one.}
\label{angdist_normalized}
\end{figure*}

For each redshift slice and for each bin (of distance between segments and angle between segments), we normalize each distribution by subtracting its mean (calculated across the parameters \mlim, \pt, \nsigtwo, and \nsigthree\ for the 2D and 3D-projected skeletons) and then dividing by its standard deviation [i.e., using the previous notation for each distribution $f(\Delta d)$ or $f(\Delta a)$ we compute $(f-\bar{f})/\sigma(f)$]. This allows us to appreciate differences in the shape of the distributions for all geometrical quantities on the same scale, regardless of the quantity which is measured. Figures \ref{distdist_normalized} and \ref{angdist_normalized} show renditions of Figs. \ref{distdist} and \ref{angdist} where the distributions have been normalized in this way. In the following we will use the terms `residual distributions' for the re-normalized distributions of Figs. \ref{distdist_normalized} and \ref{angdist_normalized} and `regular distributions' for those before renormalization of Figs. \ref{distdist} and \ref{angdist}. 

In Fig. \ref{distdist_normalized} (distance between filament segments) the fact that residual distributions are above or below zero even in the case of the best match between the 2D and 3D-projected skeletons (e.g., top leftmost panel) means that regular distributions show a deviation from the average computed across all choices of the parameters. In this case, what our algorithm aims at detecting is whether the regular distributions deviate from the average in the same way (i.e., whether the residual distributions have the same shape). In the top leftmost panel of Fig. \ref{distdist_normalized} this is expressed by the fact that the residual distributions in the 2D $\rightarrow$ 3D and 3D $\rightarrow$ 2D cases are positive or negative in the same segment distance ranges. In this case, it would mean that the regular distributions shown in Fig. \ref{distdist} have a shape that differs from the average shape across the range of values of \mlim, \pt, \nsigtwo\ and \nsigthree\ that we explore, but still overlap. On the other hand, if the residual distributions have a different shape, (e.g., because the regular distributions in Fig. \ref{distdist} deviate from the average in opposite directions, visible from the fact that when the 2D $\rightarrow$ 3D residual distribution of Fig. \ref{distdist_normalized} is positive the 3D $\rightarrow$ 2D one is negative and vice-versa), this would translate in a lack of overlap in the regular distributions of Fig. \ref{distdist}. We design our algorithm to be sensitive to this effect, which is shown in Fig. \ref{distdist_normalized} in the worst match case, where the residual distributions have very different shapes. Similar considerations apply also in the case of angle distributions.

In the case of segment angles (Fig. \ref{angdist_normalized}), the residual distributions are much more noisy (less smooth and with larger differences from bin to bin) and it is difficult to establish a trend. This translates to the fact that the distance between skeleton segments will offer a large variation of shapes in its regular distribution across the parameters explored in our analysis compared to the regular distributions of angles between filament segments. We therefore expect the distributions of distances between filament segments to yield the largest signal about how much two skeletons are similar or not. Conversely, in the case of the regular angle distributions, they are the ones closest to the average. We therefore expect these distributions to contribute less to our ability to tell two skeletons apart. We stress that the use of residual distributions and of distance and angle information combined is a new way to frame the problem of determining similarity between filament samples.

However, if we compare the residual angle distributions between the best match and the worst match scenarios, we can see how in each considered redshift slice they still have opposite slopes and different shapes in the worst match case, while they are much more similar in the best match case.

\section{Method to determine skeleton similarity}
\label{sec:distance}

We combine all the quantities described in the previous sections in a single measurement to quantify the similarity between skeletons. Our end goal is do devise a function that can associate to each 2D skeleton the most similar 3D-projected one. We employ a new measure of the quality of the match between the skeletons (2D and 3D-projected) based on several factors. We stress that in the following we make use of both regular and residual distributions. We use the measurement of the peak and width of the distance and angle distributions as performed on regular distributions (Sect. \ref{geom_prop_subsection}). We use the measurement of the differences between distribution shapes as performed on the residual distributions (Sect. \ref{diffresidualsubsection}).

\subsection{Peak and width of the regular distributions of geometrical properties}
\label{geom_prop_subsection}
In the case of the regular distance and angle distributions, the important aspects lie in the position of their peaks (for the distance distributions) and their width (for the angle distributions). Ideally, if the two skeletons considered (2D and 3D-projected) were identical, the distance distributions in the $\mathrm{3D} \rightarrow \mathrm{2D}$ and $\mathrm{2D} \rightarrow \mathrm{3D}$ cases would coincide and peak at zero. A position of the peak of the distance distribution away from zero indicates that there is a difference in the position of the filaments in the two reconstructions as the closest filament to each one of a considered skeleton is found on average at some distance. The distance in the peaks among them between the $\mathrm{3D} \rightarrow \mathrm{2D}$ and $\mathrm{2D} \rightarrow \mathrm{3D}$ cases, on the other hand, provides an indication of the amount of detail of the two skeletons. If one skeleton has significantly fewer filaments than another, the closest filament in the other skeleton will also be found at a shorter distance in the projection than in the reverse case. 

A narrow (around zero degrees) regular angle distribution indicates that the segments of filaments deemed close together in each pair of 2D and 3D skeletons are well-aligned, with minimal perpendicular angles between them. Thus, the width of the angle distribution effectively measures the degree to which filament angles deviate from alignment.

For the regular distance distributions (Fig. \ref{distdist}), a peak can be identified as the expectation value of the distribution: $\hat{X} = \sum{x \, f(x)}$ where $x$ is the value of the distance at the bin centre, $f(x)$ is the value of the distribution at the bin, and the sum is made over all the bins. We therefore define the quantities $\hat{X}_{\mathrm{3D} \rightarrow \mathrm{2D}}^{\mathrm{D}}$ and $\hat{X}_{\mathrm{2D} \rightarrow \mathrm{3D}}^{\mathrm{D}}$ for the regular distance distributions.

In the case of the regular angle distributions, we quantify the width of the distributions by means of their standard deviation, expressed as $\sigma_{X} = \sqrt{\sum{f(x)(x-\hat{X})^2}}$, where $x$ is again the value of the bin centre, $f(x)$ is the value of the distribution at the bin, the sum is made over all the bins, and $\hat{X}$ is the expectation value for the angle distributions computed as described above in the case of the distance distributions. We thus define $\sigma_{\mathrm{2D} \rightarrow \mathrm{3D}}^{\mathrm{A}}$ and $\sigma_{\mathrm{3D} \rightarrow \mathrm{2D}}^{\mathrm{A}}$ for the angle distributions.

We report the values of $\hat{X}_{\mathrm{3D} \rightarrow \mathrm{2D}}^{\mathrm{D}}$ and $\hat{X}_{\mathrm{2D} \rightarrow \mathrm{3D}}^{\mathrm{D}}$ for the regular distance distributions and the values of $\sigma_{\mathrm{2D} \rightarrow \mathrm{3D}}^{\mathrm{A}}$ and $\sigma_{\mathrm{3D} \rightarrow \mathrm{2D}}^{\mathrm{A}}$ for the regular angle distributions in Figs. \ref{distdist} and \ref{angdist} as short lines above the distributions. In particular, we can see how $\hat{X}_{\mathrm{3D} \rightarrow \mathrm{2D}}^{\mathrm{D}}$ and $\hat{X}_{\mathrm{2D} \rightarrow \mathrm{3D}}^{\mathrm{D}}$ tend to coincide in the case of the best match between skeletons and to differ in the case of the worst match. For the angle distributions, no difference can be seen in the values of $\sigma_{\mathrm{2D} \rightarrow \mathrm{3D}}^{\mathrm{A}}$ and $\sigma_{\mathrm{3D} \rightarrow \mathrm{2D}}^{\mathrm{A}}$ in the case of the considered worst matches between skeletons.

As the expectation values and the standard deviations have been derived on regular distributions before normalization, they have units of Mpc (expectation values of the regular distance distributions) and degrees (standard deviations of the regular angle distributions). These quantities are therefore non-commensurable. As our goal is to ultimately combine them in a metric to quantify the quality of the match between 2D and 3D-projected skeletons, we have to render them unit- and scale-independent through an ad-hoc normalization. We therefore define
\begin{equation}
\begin{cases}
D_{\mathrm{3D} \rightarrow \mathrm{2D}} = (\hat{X}_{\mathrm{3D} \rightarrow \mathrm{2D}}^{\mathrm{D}} - C_{\mathrm{bin}}^{\mathrm{D}})/R_{\mathrm{bin}}^{\mathrm{D}} \\
D_{\mathrm{2D} \rightarrow \mathrm{3D}} = (\hat{X}_{\mathrm{2D} \rightarrow \mathrm{3D}}^{\mathrm{D}} - C_{\mathrm{bin}}^{\mathrm{D}})/R_{\mathrm{bin}}^{\mathrm{D}} \\
A_{\mathrm{3D} \rightarrow \mathrm{2D}} = (\sigma_{\mathrm{3D} \rightarrow \mathrm{2D}}^{\mathrm{A}} - C_{\mathrm{bin}}^{\mathrm{A}})/R_{\mathrm{bin}}^{\mathrm{A}} \\
A_{\mathrm{2D} \rightarrow \mathrm{3D}} = (\sigma_{\mathrm{2D} \rightarrow \mathrm{3D}}^{\mathrm{A}} - C_{\mathrm{bin}}^{\mathrm{A}})/R_{\mathrm{bin}}^{\mathrm{A}} \\
\end{cases},
\end{equation}
where $C_{\mathrm{bin}}^{\mathrm{D}}$ and $C_{\mathrm{bin}}^{\mathrm{A}}$ are the average of the maximum value and minimum value that the distance and angle bins can take, while $R_{\mathrm{bin}}^{\mathrm{D}}$ and $R_{\mathrm{bin}}^{\mathrm{A}}$ are the differences between the maximum value and minimum value that the distance and angle bins can take. We thus obtain a `normalized' position for the peak of the segment distance regular distributions and standard deviation of the angle regular distributions, which we use in the following.

\subsection{Differences in the residual distribution shapes for geometrical properties}
\label{diffresidualsubsection}

We measure differences in the shapes of the residual distributions introduced in Sect. \ref{sec:normdist}. As such, for geometrical properties, we compute the quantities
\begin{equation}\label{ndequations}
\delta_{\mathrm{D}} = \frac{\sum{(f^{\mathrm{D}}_{\mathrm{3D} \rightarrow \mathrm{2D}}-f^{\mathrm{D}}_{\mathrm{2D} \rightarrow \mathrm{3D}})^{2}}}{N_{\mathrm{bins}}^{\mathrm{D}}}
\: \mathrm{and} \:
\delta_{\mathrm{A}} = \frac{\sum{(f^{\mathrm{A}}_{\mathrm{3D} \rightarrow \mathrm{2D}}-f^{\mathrm{A}}_{\mathrm{2D} \rightarrow \mathrm{3D}})^{2}}}{N_{\mathrm{bins}}^{\mathrm{A}}},
\end{equation}
where $f^{\mathrm{D}}$ and $f^{\mathrm{A}}$ are the residual segment distance and angle distributions, respectively, and the sum is performed across distance or angle bins. $N_{\mathrm{bins}}^{\mathrm{D}}$ and $N_{\mathrm{bins}}^{\mathrm{A}}$ are the number of bins used to derive the distributions of segment distances and angles. This latter normalization has been introduced to make sure that all quantities of Eq. \eqref{ndequations} are of the same order of magnitude and that distributions that have been derived with a larger number of bins are not artificially weighted more by having a larger number of summation terms.

\subsection{Combining distance and angle information}
\label{ddef}
We combine the shape, peak, and width of the residual and regular distance and angle distributions to better differentiate between skeletons. Using only the shape of residual segment distance and angle distributions might miss crucial details. For instance, two filament sets could have similar residual distance distribution shapes and regular distribution peaks but still be different. The peak distance value mainly reflects the relative number of large and small filaments, not their similarity. Thus, considering the peak's deviation from zero provides additional information not captured by shape alone.

We therefore add the expectation values of the regular distance distributions and the standard deviations of the regular angle distributions in quadrature to the measurement of the difference of the residual distribution shapes. We combine these measurements as
\begin{equation}
\begin{cases}
\Delta_{\mathrm{D}} = \delta_{\mathrm{D}}+D_{\mathrm{3D} \rightarrow \mathrm{2D}}^2+D_{\mathrm{2D} \rightarrow \mathrm{3D}}^2 \\
\Delta_{\mathrm{A}} = \delta_{\mathrm{A}}+A_{\mathrm{3D} \rightarrow \mathrm{2D}}^2+A_{\mathrm{2D} \rightarrow \mathrm{3D}}^2 \\
\end{cases}.
\end{equation}

We create an array of quantities $\vec{S}$, which is defined as
\begin{equation}\label{avquant_norm}
\vec{S} = (\Delta_{\mathrm{D}}, \Delta_{\mathrm{A}}).
\end{equation}
A measurement of the difference between the skeletons (i.e., of the quality of the match between a given 2D and 3D-projected skeleton) is then defined as
\begin{equation}\label{skeleton_distance_norm}
\Xi(\mathrm{2D}, \mathrm{3D}) = \sqrt{\vec{W} \cdot \vec{S}}
\end{equation}
(i.e., as the square root of the dot product between the array containing the quantities defined above, $\vec{S}$, and an array of weights, $\vec{W}$). Although the weights can theoretically have any positive real value, in the following we will only assign to them either the value of 0 or 1, in order to `switch on' or `switch off' the use of a given quantity in the definition of the measurement of the quality of the match. Assessing proper values for the weights probably requires an optimization of the skeleton difference definition and is outside the scope of this work. 

We note that this may not be the only quantitative definition of skeleton similarity achieving our goal, nor the best one. Our definition of skeleton similarity is based on how different are the distributions of filament geometrical properties introduced in Sects. \ref{sec:quantities} and \ref{sec:normdist}. Indeed several other definitions of skeleton similarity (or entirely different metrics relying on different prescriptions such as the Wasserstein metric) may be used. We postpone to a future focused paper the project of comparing different similarity measurement definitions among them and provide here the pathway to a method that can be adopted to compare skeletons among them.

Our quantitative measurement of the difference between skeletons (Eq. \ref{skeleton_distance_norm}) will be large for skeletons that are different and small for skeletons that are similar. We are thus able to minimize Eq. \eqref{skeleton_distance_norm} and to assign to each 2D skeleton its most similar 3D-projected one (i.e., the one for which the measurement of skeleton difference $\Xi$ between the chosen 2D and all possible 3D-projected skeletons is minimal). We are therefore able to define a function $\xi$ that assigns to each 2D skeleton its most similar 3D one (to which we refer in the following as $\mathrm{3D}^{\ast}$ we expect it to be unique for each 2D skeleton): $\xi(\mathrm{2D}) = \min{\Xi(\mathrm{2D}, \mathrm{3D})}\|_{\mathrm{3D}} = \mathrm{3D}^{\ast}$. In this context, the minimization is done over all possible 3D skeletons so that the function becomes a single-variable function with a 2D skeleton as input and a 3D-projected skeleton as output. This assumes that all 3D-projected skeletons are valid representations of the cosmic web.

As each 2D skeleton is defined by a set of 3 parameters (\mlim, \pt, and \nsigtwo), while each 3D-projected skeleton is defined by only one parameter (\nsigthree), $\xi(\mathrm{2D})$ is essentially a real-valued 3D function (computed for every redshift slice). The function $\xi(\mathrm{2D})$ has a complicated shape and is difficult to visualize, but it allows to define a most similar 3D-projected skeleton for each 2D one. This function can (in theory) be interpolated for values of the 2D parameters not covered here and allows us to determine, for each 2D skeleton detected with photo-$z$s, which parameters would be used to extract the skeleton in 3D if we had the possibility to extract filaments with true-$z$s in the data (and then project that skeleton from 3D to 2D). If we were able to construct expectations of the intrinsic properties of 3D skeletons for a given set of extraction parameters (e.g., by using simulations or other galaxy samples) we could use this framework to draw a connection between the filaments extracted in 2D in the EWS with photo-zs and the true properties of the skeleton. In the following, we characterize this function and its properties.

We stress that all the results presented here have been obtained using both the distance distributions and the angle distributions, which is achieved by setting $\vec{W} = (1,1)$ in Eq. \eqref{skeleton_distance_norm}. Using only distance information, or $\vec{W} = (1,0)$, would actually correspond to a quantitative version of the more qualitative approach adopted in \citet{Malavasi2017}, \citet{Laigle2018}, \citet{Kuchner2020}, and \citet{Kuchner2021}, where the distributions of segment distances among skeletons were compared in simulations. In the following we present our results in the most complete case when both distance and angle distributions are used. However, we would reach similar conclusions if only distance distributions were used, as done in previous works.

\section{Characterizing skeleton similarity}
\label{sec:characterizing}
As we defined a function that provides for each 2D skeleton its most similar 3D-projected one ($\mathrm{3D}^{\ast}$), we are able to perform a comparison between the properties of a given 2D skeleton and its best matching 3D one. To achieve this, we use independent skeleton properties not previously measured for similarity, but crucial for the scientific analysis. Specifically, we incorporate \emph{astrophysical properties}, which relate to the properties of galaxies in the cosmic web and the connectivity--mass relation.

\subsection{Mass gradient with distance from filament axis.}
Motivated by the relation between the distance of a galaxy from a filament spine and its mass, we measure the distance of each galaxy to the closest segment of the closest filament (\dfil). We then measure the average stellar mass of galaxies in bins of \dfil. We refer to this measurement as "mass gradient" in the following: $M(d_{\mathrm{fil}})_{\mathrm{3D}}$ and $M(d_{\mathrm{fil}})_{\mathrm{2D}}$ (for the 3D-projected and 2D skeletons, respectively). The distance \dfil~is measured in 2D also for the 3D-projected skeleton. We exclude galaxies close to the nodes to prevent picking up the mass gradient related to clusters in the following way. We derive for each galaxy the distance to the densest node following the path of the closest filament. We do this by measuring the length of the portion of the filament closest to the galaxy that connects the projected position of the galaxy onto that filament and the densest node at the filament extremity (\dskel). We also derive the Euclidian distance to the closest node (\dnode). We then perform a measurement of $M(d_{\mathrm{fil}})_{\mathrm{3D}}$ and $M(d_{\mathrm{fil}})_{\mathrm{2D}}$ only for galaxies with \dskel~and \dnode~both larger than the 25th percentiles of the \dskel~and \dnode~distributions.\footnote{Typically, the median values of the 25th percentiles of the \dskel~and \dnode~distributions over the considered redshift interval are between 0.7 Mpc and 4 Mpc for all skeleton realizations} A visual representation of the distances \dfil, \dskel, and \dnode\ can be found in Fig. 1 of \citet{Malavasi2022spin}.

\begin{figure*}
\centering
\includegraphics[width = \textwidth, trim = 1.5cm 1cm 1.5cm 1cm, clip=true]{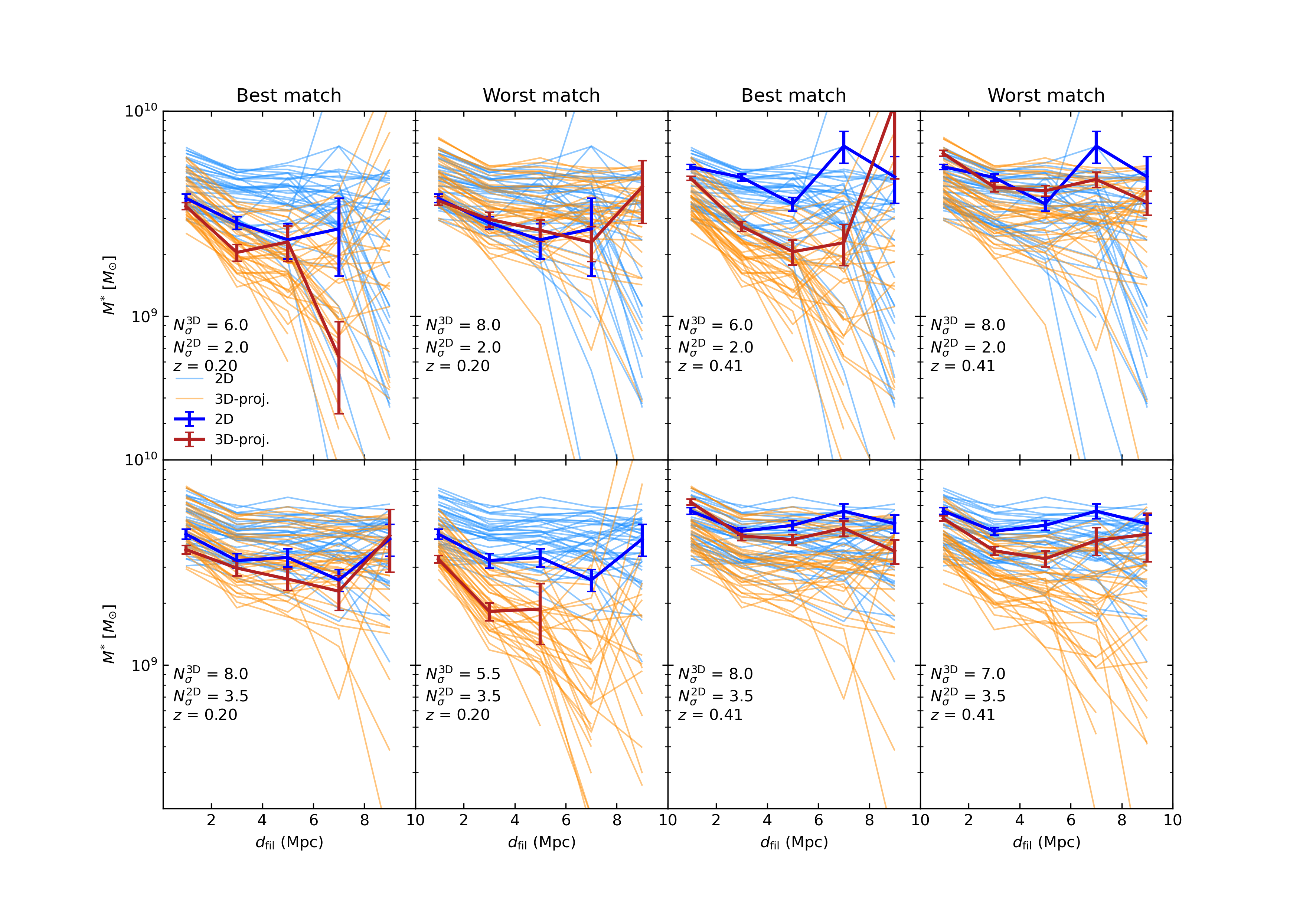}
\caption{Same as Fig. \ref{distdist}, but for the average galaxy stellar mass in bins of distance from the spine of the filaments.}
\label{massgrad}
\end{figure*}

In Fig. \ref{massgrad} we show the mass gradient as measured using 2D (blue) and 3D-projected skeletons (red) in the case of the best and the worst match between 2D and 3D-projected skeletons. The worst matching 3D-projected skeleton for a given 2D one is found using the opposite approach of maximizing $\Xi$ by computing $\max{\Xi(\mathrm{2D}, \mathrm{3D})}\|_{\mathrm{3D}}$. As done previously, we only report the intermediate case of galaxy mass limit $\logten($\mlim$/M_{\sun}) = 10.25$ and \pt$\,= 0.4$. For two different redshift slices, we report the case of the best and the worst match between skeletons, for a low and a high value of \nsigtwo.

In the case of 2D skeletons, we always recover a mass gradient which has the same shape or shallower as the one of its best-matching 3D-projected skeleton. This was confirmed by measuring the quantity $[M(d_{\mathrm{fil}})_{\mathrm{3D}^{\ast}} - M(d_{\mathrm{fil}})_{\mathrm{2D}}]/M(d_{\mathrm{fil}})_{\mathrm{3D}^{\ast}}$ for all redshifts and all combinations of parameters \mlim, \pt, and \nsigtwo\ (not shown here).

If we compare the mass gradient recovered for a high-persistence 2D skeleton with the one recovered for its worst-match 3D-projected skeleton, the two have clearly different shapes, while in the case of a low-persistence 2D skeleton the two look similar.

\subsection{Connectivity-mass and Multiplicity-mass relations.}
Connectivity refers to the count of filaments connected to a maximum (e.g., galaxy cluster) which in turn connect to saddle points, excluding any branching points (bifurcations). If bifurcations are incorporated in the skeleton, we measure multiplicity instead. Both connectivity and multiplicity of a node correlate with the mass of its associated structure. More massive nodes are connected to more filaments.

We use both the connectivity-mass, $\kappa(M)_{\mathrm{3D}}$ and $\kappa(M)_{\mathrm{2D}}$ (for the 3D-projected and 2D skeletons, respectively) and the multiplicity-mass relations, $\mu(M)_{\mathrm{3D}}$ and $\mu(M)_{\mathrm{2D}}$ (for the 3D-projected and 2D skeletons, respectively) as part of our assessment of the \disperse\ algorithm for reconstructing 2D skeletons using photo-$z$s.

The initial step involves assigning a mass to each node by finding the structure associated to that node (which is in turn associated to a critical point extracted by \disperse, either a maximum or bifurcation). Since \disperse~operates independently of scale, determining the scale at which to measure connectivity-mass and multiplicity-mass relations is essential. To address this, we attach mass measurements to nodes, either typical of clusters or galaxies, enabling analysis at both large and small scales. One method involves associating each critical point with the virial mass of its closest central halo (or the central halo of its closest galaxy), potentially representing the scale of groups or clusters. Alternatively, we attribute the stellar mass of the nearest central galaxy to each critical point, interpreting them to be at the scale of galaxies.

However, a challenge arises due to projection effects, particularly in 2D mass association, where galaxies or haloes may appear closer to critical points on the plane of the sky than they are along the line of sight. To mitigate this, we implement mass ranking, associating each critical point with the mass of the most massive halo or central galaxy within a given radius.

Connectivity and multiplicity are then measured, with connectivity assessed only for maxima (while we use for the measurement of multiplicity both maxima and bifurcations). Both measurements are taken at a distance from the chosen critical point (maxima or bifurcations) to simulate observational scenarios where critical points must be matched to clusters or galaxies before measuring connectivity or multiplicity. We use distance thresholds corresponding to cluster and galaxy scales, summarizing our findings in Table \ref{connsummary}.

\begin{table*}
\caption{Summary of connectivity and multiplicity measurements}
\label{connsummary}
\centering
\begin{tabular}{c c c}
\hline\hline
Method & Cluster scale & Galaxy scale \\ 
\hline
Distance of $\kappa$ and $\mu$ measurement & $R_{\mathrm{vir}}$ & 1.5 Mpc \\
\hline
Mass of CP (proximity) & $M_{\mathrm{vir}}$ of host halo of closest halo to CP & Stellar mass of closest central galaxy to CP \\
\hline
Mass of CP (mass ranking) & \makecell{Max $M_{\mathrm{vir}}$ of host halo of sub-haloes \\ having CP within sphere \\ of radius $R_{\mathrm{vir}}$ of host halo} & \makecell{Stellar mass of most massive central galaxy \\ among those with CP within 1.5 Mpc} \\
\hline
\end{tabular}
\end{table*}

The virial radius for each central halo is computed using the value of its virial mass as
\begin{equation}\label{mvirtorvir}
R_{\mathrm{vir}} = \sqrt[3]{\frac{3M_{\mathrm{vir}}}{4\pi \, \Delta_{\mathrm{vir}}\rho_{\mathrm{c}}(z)}}, 
\end{equation}
where $\rho_{\mathrm{c}}(z)$ is the critical density of the Universe at the redshift of the considered halo of mass $M_{\mathrm{vir}}$. We use an overdensity $\Delta_{\mathrm{vir}} = 200$ for simplicity (although we are aware that at our redshifts of interest other values may be more indicated, see e.g. \citealt{Bryan1998}).

The connectivity-mass and multiplicity-mass relations are derived as the mean value of the connectivity or multiplicity in bins of $\mathrm{M}_{\mathrm{vir}}$ or stellar mass. We note that our measurement of connectivity (where we measure it for all maxima in the recovered skeleton) is different than what done in other works \citep[e.g.][]{Sarron2019, DarraghFord2019}, where connectivity is measured only in galaxy clusters of a given sample.

\begin{figure*}
\centering
\includegraphics[width = \textwidth, trim = 1.5cm 1cm 1.5cm 1cm, clip=true]{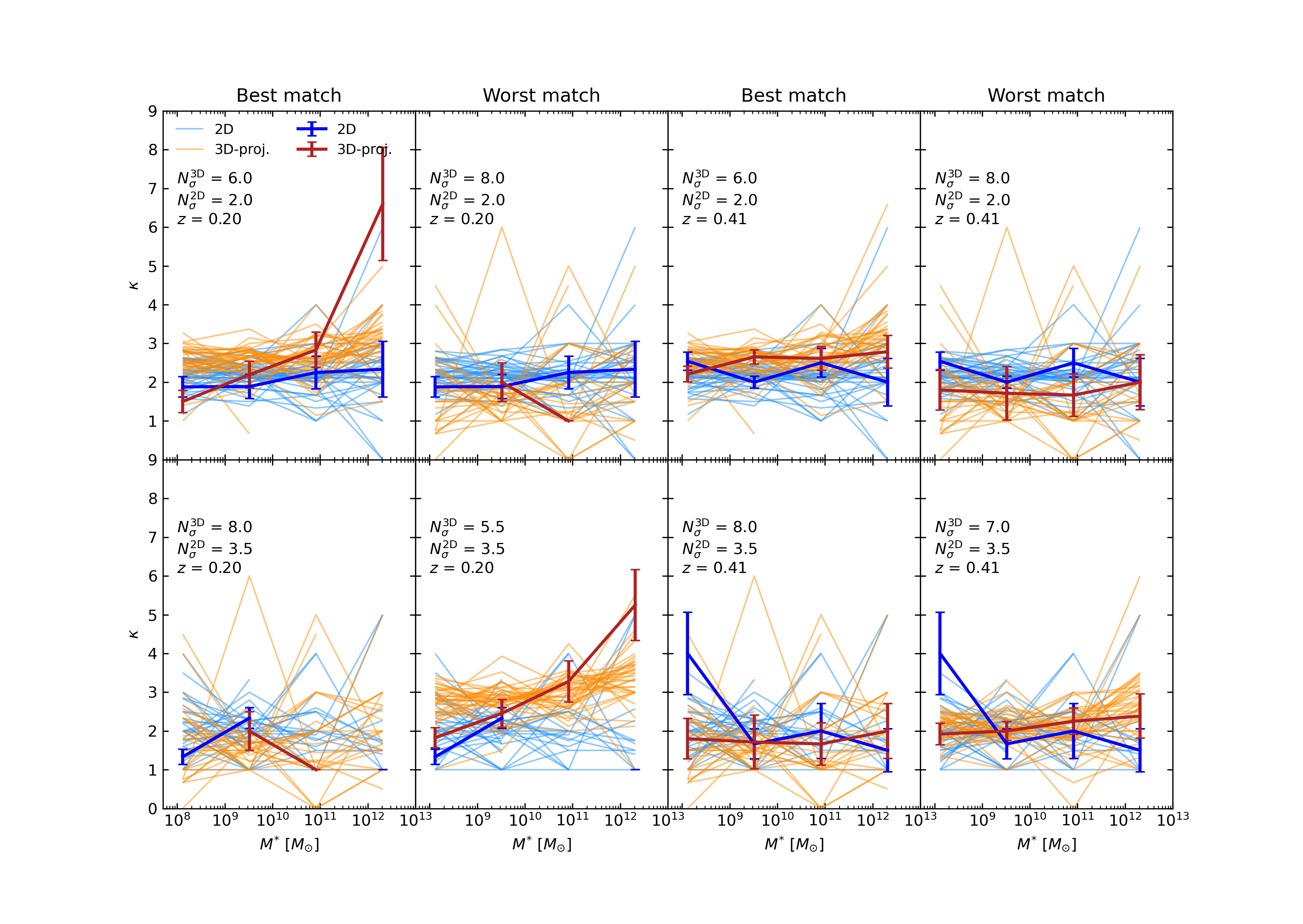}
\caption{Same as Fig. \ref{distdist}, but for the average connectivity in bins of stellar mass.}
\label{connmass_galaxy}
\end{figure*}

Figure \ref{connmass_galaxy} shows the connectivity--mass relation for critical points when considering the galaxy scale, with galaxies matched to critical points based on proximity. As before we only report the case of galaxy mass limit $\logten($\mlim$/M_{\sun}) = 10.25$ and \pt$\,= 0.4$ for two relevant redshift slices and for the case of the best and worst match between a high-persistence (low-persistence) 2D skeleton and its corresponding 3D-projected one.

The measured connectivity--mass trends are difficult to establish already for the 3D-projected skeleton case. While for some parameter combinations a weak, but clear trend can be recovered for the 3D-projected skeleton, in other cases no trend is visible, with connectivity having no dependence on stellar mass. In the most extreme cases, some mass bins are not populated by any critical point, as all masses associated to the critical points of the skeleton were at the high end or at the lower end of the available mass range.  

The challenge of capturing the relation is valuable information for our goal: it teaches us that specific combinations of \disperse~parameters should not be considered for connectivity--mass relation studies. Furthermore, we never recover the connectivity--mass relation in the 2D skeleton, regardless of the parameter combination.

As for the case of the mass gradient, we further confirmed this by measuring the quantity $[\kappa(M)_{\mathrm{3D}^{\ast}} - \kappa(M)_{\mathrm{2D}}]/\kappa(M)_{\mathrm{3D}^{\ast}}$ for all redshifts and all combinations of parameters \mlim, \pt, and \nsigtwo\ (not shown here). Similarly, we never detect the multiplicity--mass relation in the 3D-projected skeleton, regardless of the parameter combinations. As a consequence, we do not use the multiplicity--mass relation in evaluating the skeleton's recovery accuracy with \Euclid-like photo-$z$s and advise others considering its use for similar assessments to perform more specific tests exploring this quantity in greater detail.

\subsection{Trends in the persistence thresholds for best-matching skeletons}
\begin{figure*}
\centering
\includegraphics[scale=0.84, trim = 1cm 2.25cm 1cm 0cm, clip=true]{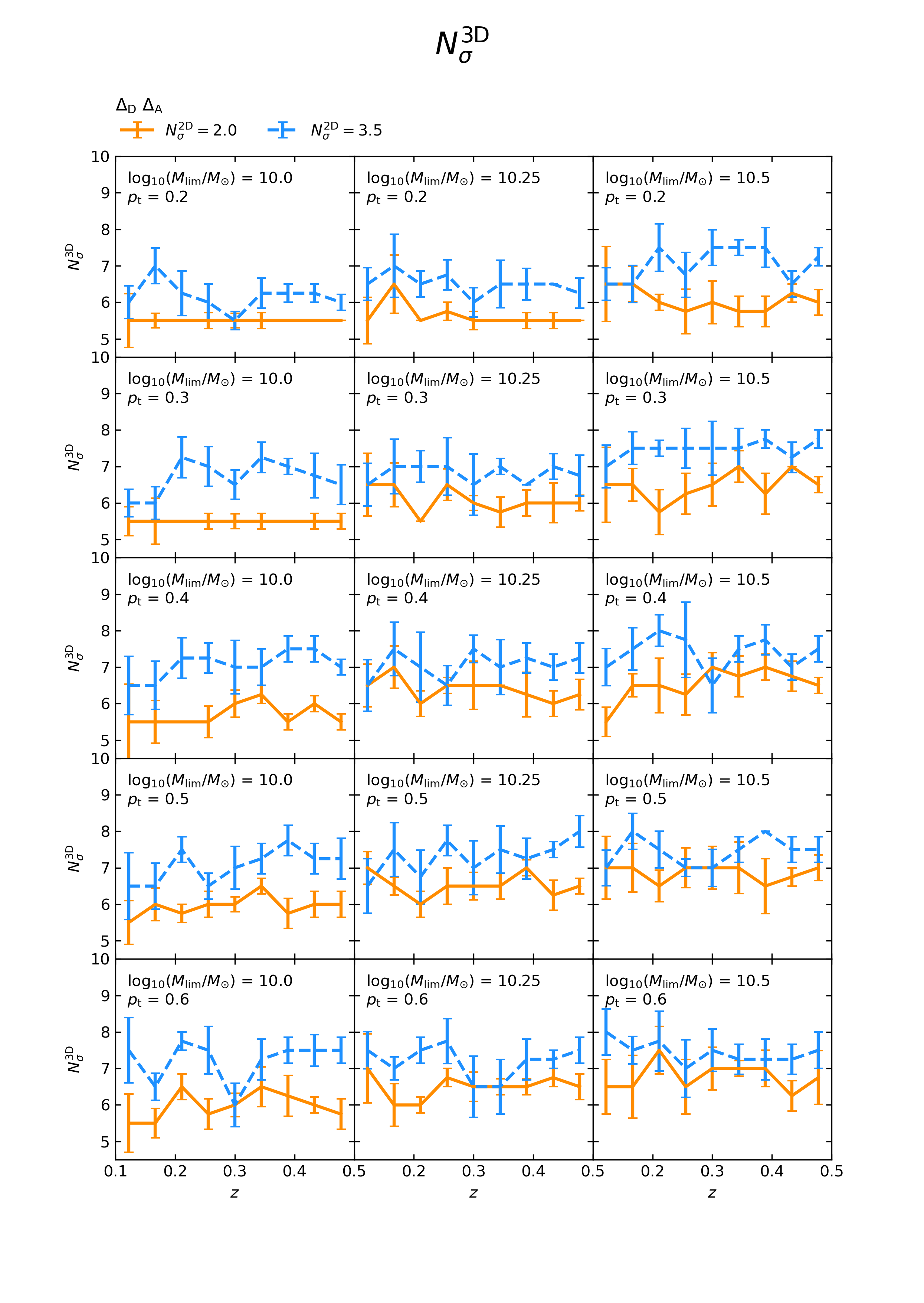}
\caption{Parameter \nsigthree\ (persistence threshold) of the skeleton $\mathrm{3D}^{\ast}$ for each 2D skeleton explored in this work as a function of redshift. Weight vector is set to $\vec{W} = (1,1)$. Parameter \mlim\ increases along columns, parameter \pt\ increases along rows. In each panel, orange solid line refers to \nsigtwo$ = 2.0$, cyan dashed line refers to \nsigtwo$ = 3.5$. The distributions have been re-binned in redshift to enhance trends. Errorbars are the standard deviation of the distribution of \nsigthree\ within each large redshift bin.}
\label{soutsin}
\end{figure*}

Next, we identify the value of the important persistent threshold parameter \nsigthree\ (see Sect. \ref{sec:fil_extraction}; the variable parameter in \disperse\ regulating signal-to-noise) of the skeleton $\mathrm{3D}^{\ast}$ as a function of redshift for different combinations of the parameters of the 2D input skeleton. This is the most direct way in which our algorithm that quantifies similarity between skeletons is able to allow us to evaluate the performance of filament extraction using photo-$z$s in the EWS. By determining the persistence threshold of the best-matching 3D-projected skeleton as a function of the input 2D one we are able to establish to a degree what type of 3D skeleton we would be observing if we had access to galaxies true-$z$s given an extraction performed in 2D with photo-$z$s. \nsigthree\ of the skeleton $\mathrm{3D}^{\ast}$ as a function of redshift for different combinations of the parameters of the 2D input skeleton is shown in Fig. \ref{soutsin}. In order to enhance trends with redshift we have re-binned our distributions in bins of width $\Delta z = 0.045$. We present the original non-rebinned version of this image in Appendix \ref{nonrebinned}. Figure \ref{soutsin} shows \nsigthree of the skeleton $\mathrm{3D}^{\ast}$ for two extreme cases in the choice of \nsigtwo. Other choices of this parameter lie in between the lines and represent intermediate possibilities. Had we used only the information from the segment distance distributions in the Fig. \ref{soutsin}, our conclusions would not have changed.

This figure shows essentially three trends. First of all, \nsigthree\ of the $\mathrm{3D}^{\ast}$ skeleton increases as a function of \nsigtwo\ at all redshifts. This trend is expected and indicates that as we focus on only the most persistent structures in the 2D skeleton, we find a correspondence in the 3D-projected skeletons that also increase their persistence threshold.

Second and third, \nsigthree\ of the $\mathrm{3D}^{\ast}$ skeleton increases with \pt\ (at fixed \mlim) and with \mlim\ (at fixed \pt, with this trend being the strongest between the two). This is also expected as overall an increase in both \mlim\ and \pt\ parameters corresponds to a reduction in the number of galaxies available as tracers for the extraction of the cosmic web (see, e.g., \citealt{Bermejo2024}). Most importantly, removing lower mass galaxies (while it lowers the number of galaxies with larger uncertainties in their photo-$z$) has the stronger effect of reducing the overall amount of tracers which are located in the structures that we aim at recovering (as highest mass galaxies are located preferentially in nodes).

\begin{figure*}
\centering
\includegraphics[scale=0.84, trim = 1cm 2.25cm 1cm 0cm, clip=true]{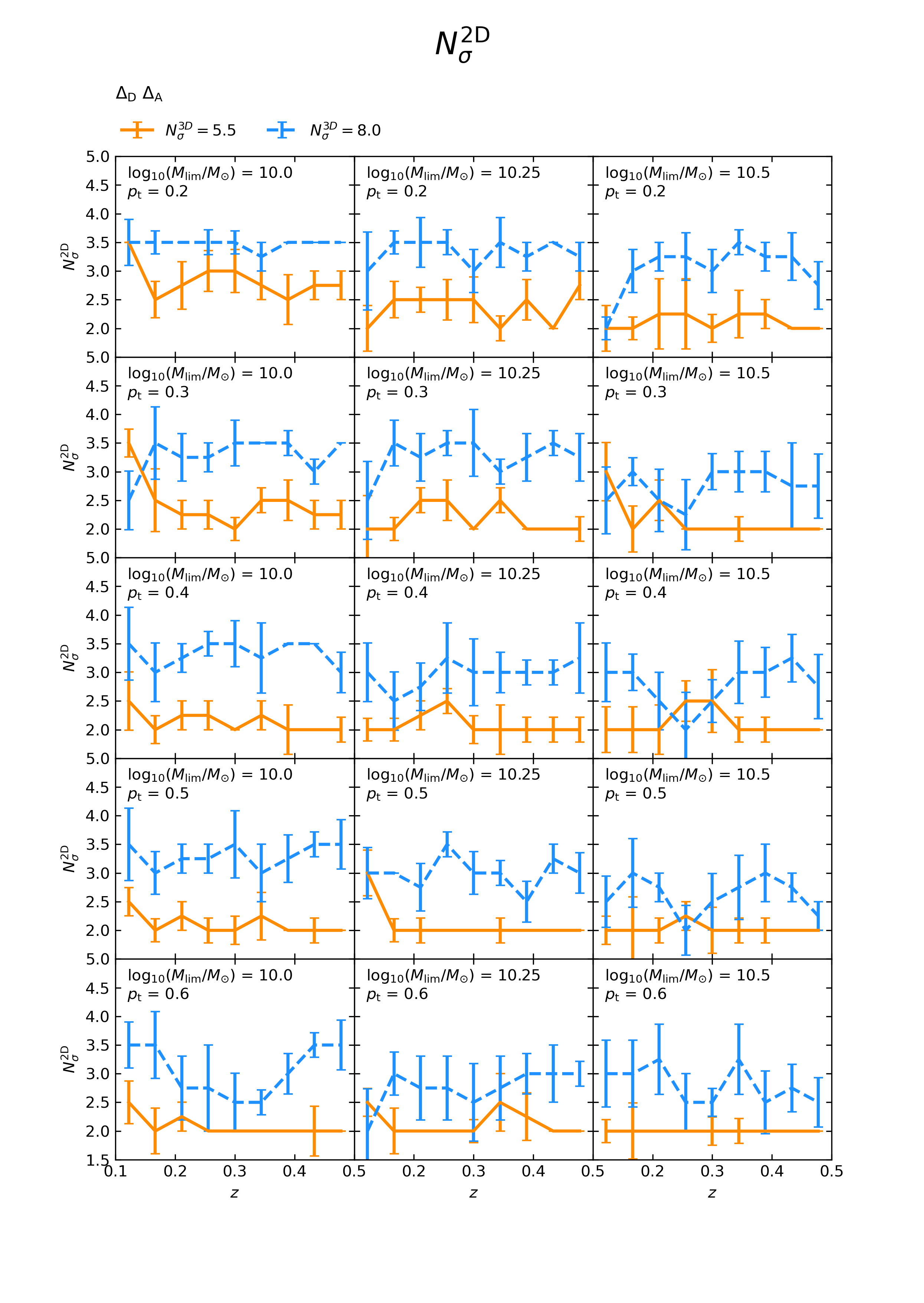}
\caption{Same as Fig. \ref{soutsin}, but showing the parameter \nsigtwo\ (persistence threshold) of the 2D projected skeleton that has to be chosen to keep the persistence threshold of the best-match 3D-projected skeleton constant as a function of redshift. In each panel, orange solid line refers to \nsigthree$ = 5.5$, cyan dashed line refers to \nsigtwo$ = 8$.}
\label{sinfixsout}
\end{figure*}

We also used our measurement of skeleton similarity to identify the value of \nsigtwo\ for the 2D skeleton in order to keep the \nsigthree\ of the best-matching 3D one constant. This is also another direct way to quantify the performance of filament extraction with photo-$z$s in the EWS. In fact, it provides an indication, to a degree, of what value to adopt for the tunable \disperse\ parameter \nsigtwo\ in the 2D extraction performed with photo-$z$s that would allow us to achieve a desired quality in the filament reconstruction in 3D had we access to galaxies' true-$z$s. The value of \nsigtwo\ for the 2D skeleton to use in order to keep the \nsigthree\ of the best-matching 3D one constant has been computed by minimizing Eq. \eqref{skeleton_distance_norm} with respect to the 2D skeleton by computing $\zeta(3D) = \min{\Xi(\mathrm{2D}, \mathrm{3D})}\|_{\mathrm{2D}}$, compare with the definition of $\xi(2D)$ in Sect. \ref{ddef}. This is shown in Fig. \ref{sinfixsout}. 

This figure shows that, as expected from the previous analysis, there is a strong correlation between \nsigtwo\ and \nsigthree. In particular, for all \mlim\ and \pt\ combinations (except the most extreme ones), the highest persistence 3D-projected skeleton is obtained when \nsigtwo\ is above 3, while for the lowest persistence 3D-projected skeleton, \nsigtwo\ has to be between 2 and 2.5. This figure shows the versatility of our method for measuring the similarity of 2D and 3D-projected skeletons and its application to the specific range of parameters explored in the case of the EWS.

\section{Discussion and conclusions} 
\label{sec:conclusions}
Assessing the similarity between sets of filaments is not an easy task. The main result of this work is the creation of a method to efficiently compare sets of filaments derived with \disperse\ on similar data sets. The derivation of this method is performed starting from geometrical and astrophysical principles which focus on the typical use cases that require filament samples: comparison of the cosmic web structure (geometrical properties), impact of the cosmic web on galaxy and cluster evolution (mass gradient and connectivity--mass relation).

Based on our approach we determine that the filaments detected with \disperse\ using photo-$z$s in \Euclid can be related to reference skeletons built with true-$z$s. In particular, the correlation between \nsigtwo\ and \nsigthree\ (i.e., the main \disperse\ parameter that establishes the degree of reliability of the filaments that are included in the extracted sample) indicates that the robustness (persistence) of structures extracted using photo-$z$s is correlated with the robustness of structures extracted using true-$z$s.

The trends with \pt\ and \mlim\ suggest that the number of tracers used has a greater impact on our reconstruction than photo-$z$ uncertainty (for the range of photo-$z$ uncertainties explored here, it is unclear whether this trend would hold also when photo-$z$s with smaller uncertainties were considered). Including more galaxies, even with larger photo-$z$ uncertainty, by using less conservative \pt\ and \mlim\ cuts, enhances our ability to recover the relationships between 2D skeletons and their 3D projections.

Specifically, skeletons built with photo-$z$s \textit{can} be used to investigate properties of galaxies near the cosmic web, such as how galaxy stellar mass depends on the distance from the filament axis. Although the steepness of the mass gradient recovered with the 2D skeleton will be less pronounced than in the real case, it will still be detectable.

Finally, the connectivity--mass relation cannot be unequivocally detected even in our reference 3D-projected case. As we are unable to use it to assess skeleton similarity, it is likely that more in-depth analyses, focusing specifically on the measurement of connectivity, are needed before it can be studied within the context of the EWS. The application of the results derived here to skeletons extracted in 2D from real EWS data will be possible, provided that a comparison skeleton extracted in 3D and projected is available in an overlapping area (see, e.g., Kraljic et al., in preparation) and that further investigations on, e.g., the impact of masks on skeleton extraction and on how to properly measure connectivity is executed.

This work significantly improves and expands upon previous efforts to compare filament catalogues. Future works which will employ other methods (e.g., information theory or machine learning) will further expand our capability to compare sets of structures marked by a complicated, connected, anisotropic, and multiscale nature.

\begin{acknowledgements} 
\AckEC

NM acknowledges funding by the European Union through a Marie Sk{\l}odowska-Curie Action Postdoctoral Fellowship (Grant Agreement: 101061448, project: MEMORY). Views and opinions expressed are however those of the author only and do not necessarily reflect those of the European Union or of the Research Executive Agency. Neither the European Union nor the granting authority can be held responsible for them. UK acknowledges financial support from the UK Science and Technology Facilities Council (STFC; grant ref: ST/T000171/1). This work was supported by the project "Galaxy evolution in the cosmic web" of Swiss National Science Foundation.

This work has made use of CosmoHub. CosmoHub has been developed by the Port d'Informaci{\'o} Cient{\'i}fica (PIC), maintained through a collaboration of the Institut de F{\'i}sica d'Altes Energies (IFAE) and the Centro de Investigaciones Energ{\'e}ticas, Medioambientales y Tecnol{\'o}gicas (CIEMAT) and the Institute of Space Sciences (CSIC \& IEEC). CosmoHub was partially funded by the ``Plan Estatal de Investigaci{\'o}n Científica y T{\'e}cnica y de Innovaci{\'o}n'' program of the Spanish government, has been supported by the call for grants for Scientific and Technical Equipment 2021 of the State Program for Knowledge Generation and Scientific and Technological Strengthening of the R+D+i System, financed by MCIN/AEI/ 10.13039/501100011033 and the EU NextGeneration/PRTR (Hadoop Cluster for the comprehensive management of massive scientific data, reference EQC2021-007479-P) and by MICIIN with funding from European Union NextGenerationEU (PRTR-C17.I1) and by Generalitat de Catalunya.

\end{acknowledgements}

\bibliographystyle{aa.bst}
\bibliography{biblio_2d}


\newpage
\appendix
\onecolumn

\section{Trends using distributions not rebinned in redshift}
\label{nonrebinned}
\begin{figure*}
\centering
\includegraphics[scale=0.77, trim = 1cm 2.25cm 1cm 0cm, clip=true]{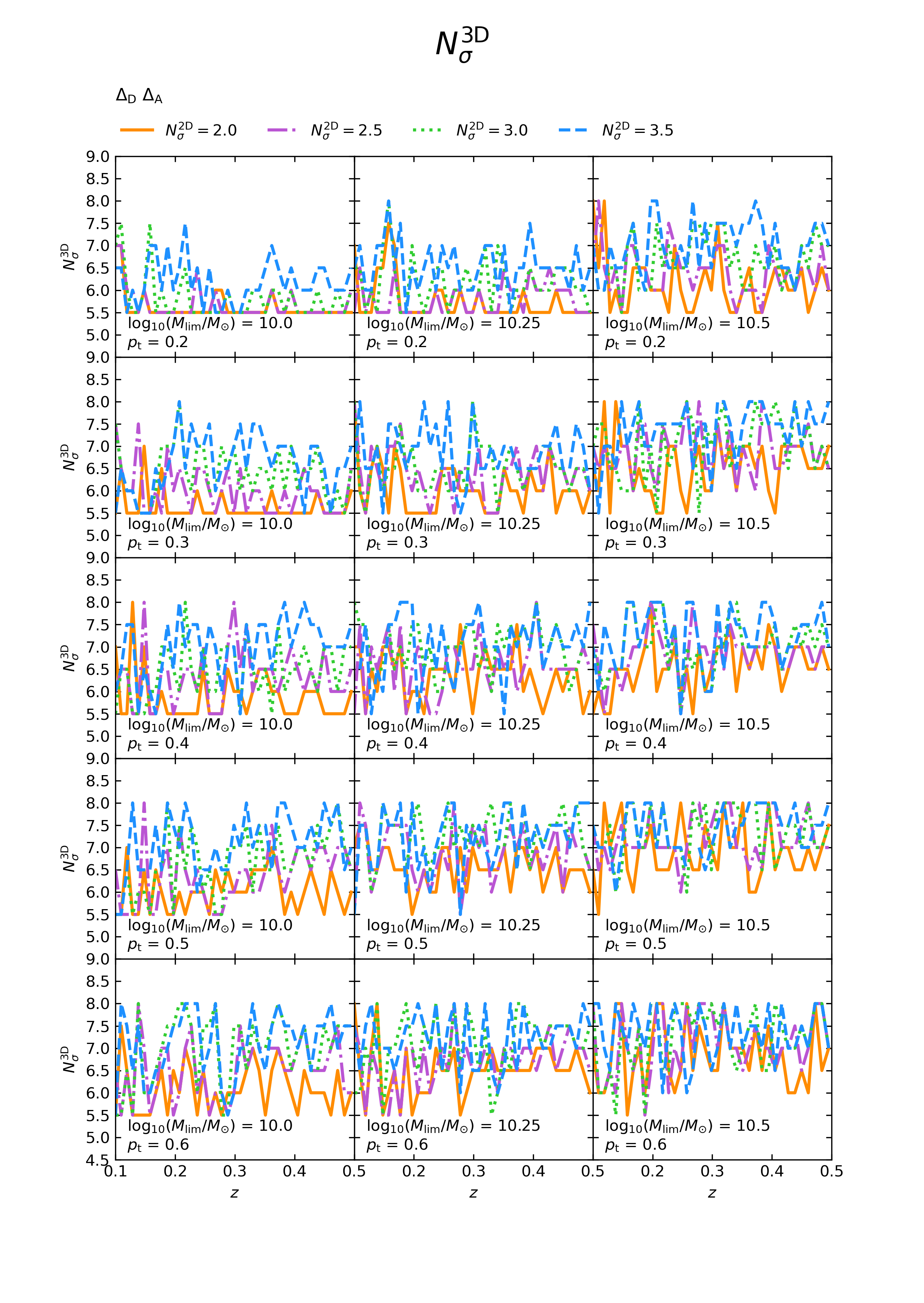}
\caption{Same as Fig. \ref{soutsin}, but with the distributions not rebinned in redshift. In each panel, the orange solid line refers to \nsigtwo$ = 2.0$, the purple dash-dotted line refers to \nsigtwo$ = 2.5$, the green dotted line refers to \nsigtwo$ = 3.0$, and the cyan dashed line refers to \nsigtwo$ = 3.5$.}
\label{soutsin_nonreb}
\end{figure*}

We show here a rendition of Fig. \ref{soutsin} without the rebinning in redshift (Fig. \ref{soutsin_nonreb}). This figure shows that the trends identified in Sect. \ref{sec:characterizing} are present also in the non-rebinned case, albeit less evident. We also include the intermediate \nsigtwo values that were excluded from Fig. \ref{soutsin} for completeness.

\clearpage
\onecolumn
\section{Use of the Flagship simulation}
\label{flagship}
The Flagship Mock Galaxy Catalogue version 2.1.10\footnote{Available to the \Euclid Consortium on CosmoHub \citep{Carretero2017, Tallada2020}: \url{https://cosmohub.pic.es/catalogs/298}} \citep{Castander2024} is based on the Flagship2 N-body simulation, which tracks the evolution of $4 \times 10^{12}$ dark matter particles in a box of $3600 h^{-1} \mathrm{Mpc}$ on a side, for a particle mass resolution of $10^9 h^{-1} M_{\sun}$. It was performed using values for the cosmological parameters of $h = 0.67$, $\Omega_{\mathrm{m}} = 0.319$, $\mathrm{A}_{\mathrm{s}} = 2.1 \times 10^{-9}$, and $n_{\mathrm{s}} = 0.96$. Haloes in the simulation were identified with the \texttt{ROCKSTAR} halo finder \citep{Behroozi2013}.

The lightcone covers $\sim 5000 \deg^2$ on the plane of the sky in the redshift range $z \in [0,3]$, from which we extracted an area of $\sim 50 \deg^2$ to perform our analysis. The lightcone itself contains $3.3 \times 10^9$ galaxies with magnitude $H \leq 26$. It provides galaxy positions, true redshifts, galaxy spectroscopic and photometric information as well as additional properties such as lensing information, shape parameters and central/satellite classification. Photo-$z$s were computed for the galaxies using NNPZ, (\citealt{Cunha2009}, see also \citealt{Desprez2020}).

\begin{figure*}
\centering
\includegraphics[width = \textwidth, trim = 1.5cm 1cm 1.5cm 1cm, clip=true]{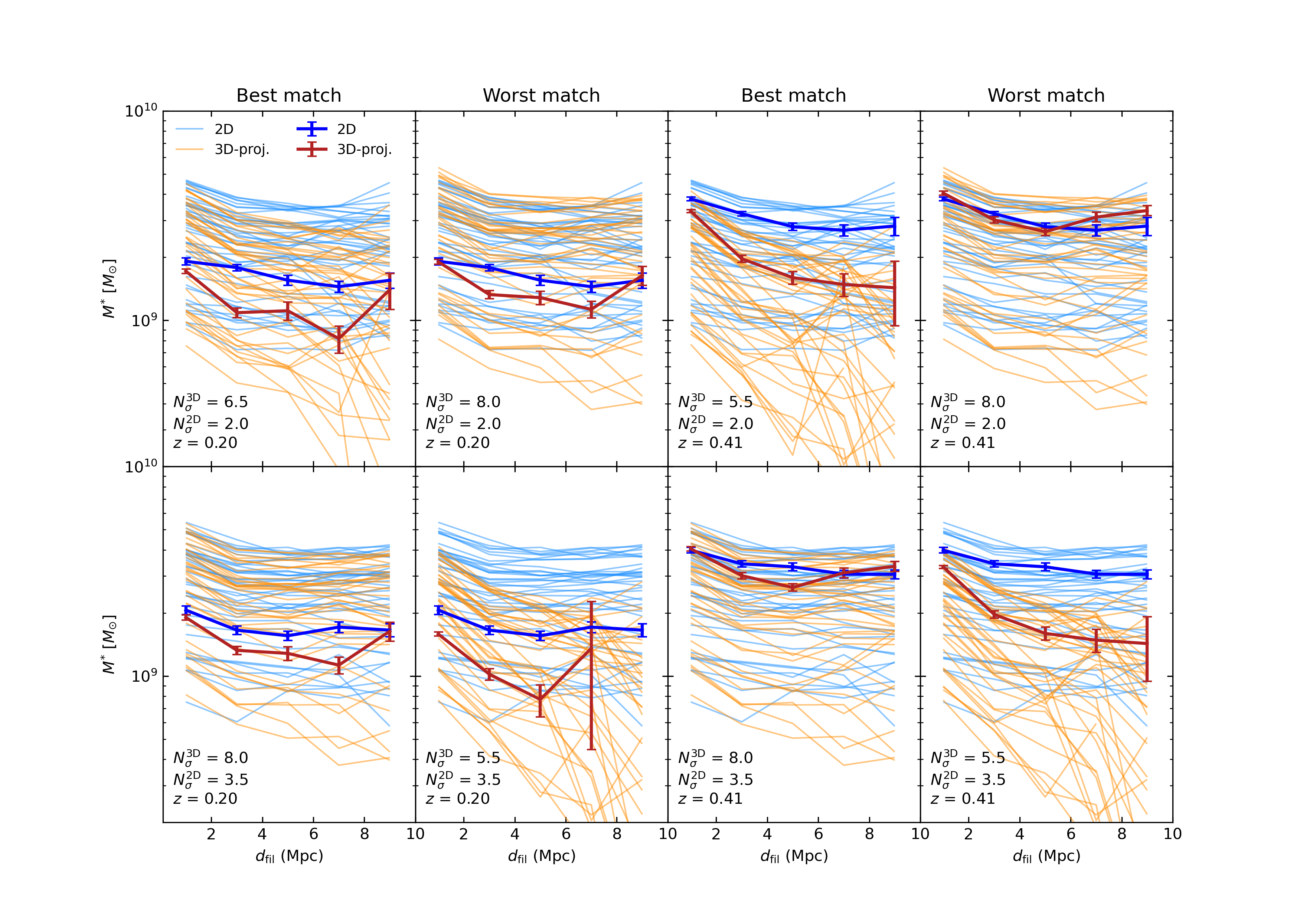}
\caption{Same as Fig. \ref{massgrad}, but for the Flagship Mock Galaxy Catalogue.}
\label{massgrad_flagship}
\end{figure*}

We repeat our analysis using the Flagship Mock Galaxy Catalogue, to check that our results are consistent also when using different simulated data. Specifically, Fig. \ref{massgrad_flagship} displays the mass gradient derived from both high-persistence and low-persistence 2D skeletons, using the parameter combination $\logten($\mlim$/M_{\sun}) = 10.25$ and \pt$\,= 0.4$, across two pertinent redshift slices.\footnote{Note that due to the fact that the cosmology used to create the Flagship Mock Galaxy Catalogue is different than the one used for GAEA, the number of redshift slices in which we divide our simulated galaxy sample in this case is 41 instead of 39.} This is compared against the corresponding best-match and worst-match 3D-projected skeletons (refer to Fig. \ref{massgrad} in the case of the GAEA mock galaxy catalogue).

Similar to the GAEA case, we observe that the mass gradient recovered from the 2D skeleton can match or be shallower compared to its best-match 3D-projected skeleton. This agreement is particularly notable for high-redshift skeletons, where the mass gradients align in the case of a well-matched pair and diverge in the case of a poor match.

\begin{figure*}[h!]
\centering
\includegraphics[scale=0.84, trim = 1cm 2.25cm 1cm 0cm, clip=true]{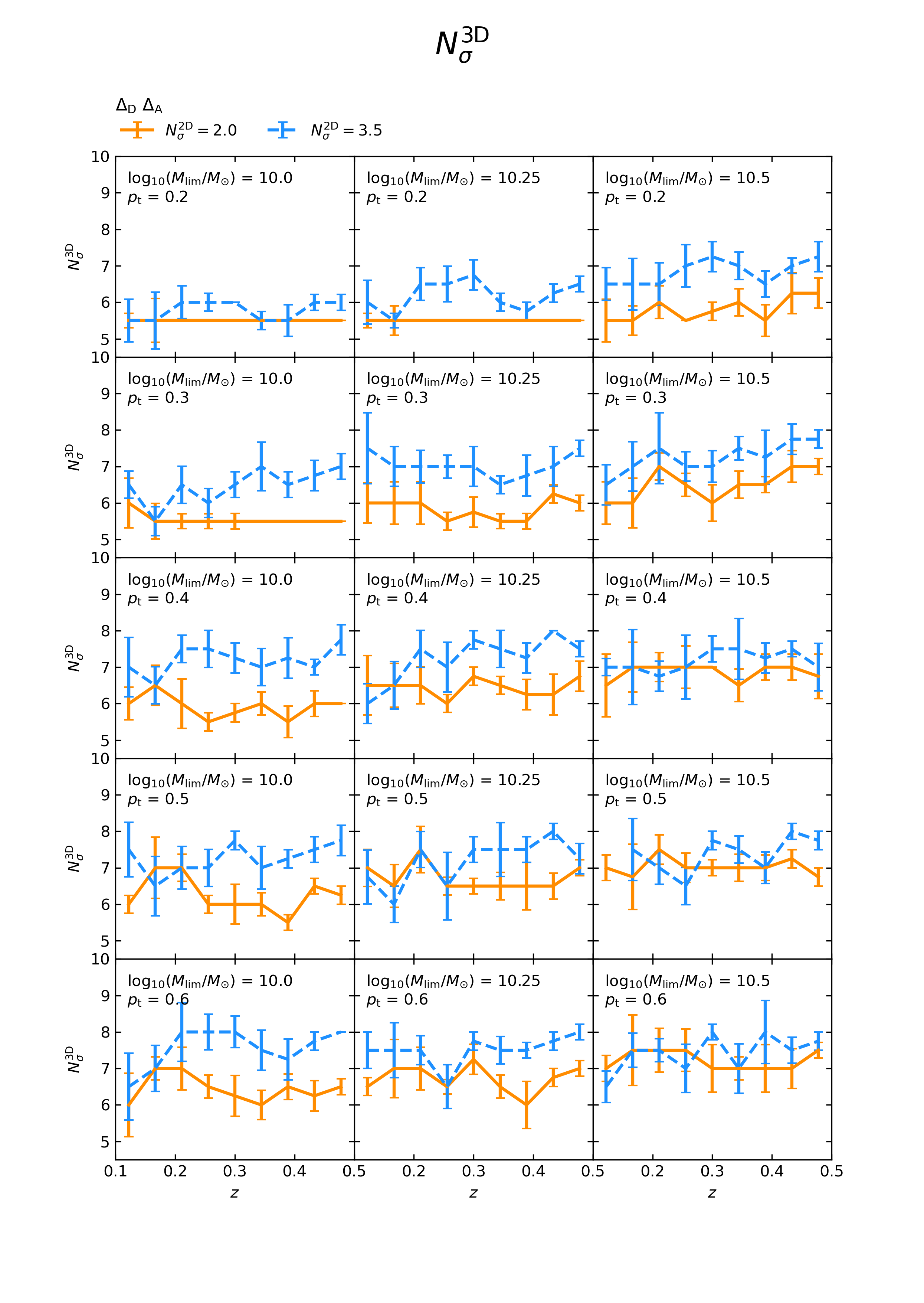}
\caption{Same as Fig. \ref{soutsin}, but for the Flagship Mock Galaxy Catalogue.}
\label{soutsin_flagship}
\end{figure*}

Figure \ref{soutsin_flagship} shows a rendition of Fig. \ref{soutsin} obtained using the Flagship Mock Galaxy Catalogue. These simulated data were processed in the same way as GAEA, following the method detailed in the main body of the paper. This figure shows that the trends we identified using the GAEA ECLQ lightcone are recovered also when using Flagship and that the conclusions we reached are preserved. In particular, \nsigthree\ of the $\mathrm{3D}^{\ast}$ skeleton still increases with both \mlim\ and \pt, while also showing a strong relation to \nsigtwo\ of the input skeleton.

\end{document}